\documentclass[10pt,twocolumn,a4aper,conference]{IEEEtran}
\usepackage[english]{babel}
\usepackage[latin1]{inputenc}
\usepackage{amsmath, amssymb}
\usepackage{pstricks, pst-node, pstricks-add}
\usepackage{subfigure}
\usepackage{bm}
\usepackage{multirow}
\usepackage{rotating}
\usepackage{stmaryrd}
\usepackage{units}
\usepackage{rotating}
\usepackage{graphics}
\usepackage[printonlyused]{acronym}	    

\newcommand{\etal}[0]{\textit{et al.}}

\newcommand{\DefSet}[1]{\mathbb{#1}}
\newcommand{\Matrix}[1]{\mathbf{#1}}
\newcommand{\MatrixTime}[2]{\Matrix{#1}^{[#2]}}

\newrgbcolor{verylightgray}{0.8 0.8 0.8}
\newrgbcolor{verylightblue}{0.8 0.8 1.0}
\newrgbcolor{verylightred}{1.0 0.8 0.8}
\newrgbcolor{lightblue}{0.7 0.7 0.9}
\newrgbcolor{lightgreen}{0.7 0.9 0.7}
\newrgbcolor{lightred}{0.9 0.7 0.7}
\newrgbcolor{darkgreen}{0.2 0.6 0.2}
\newrgbcolor{darkred}{0.6 0.2 0.2}
\newrgbcolor{DLcolor}{0.6 0.1 0.1}
\newrgbcolor{ULcolor}{0.1 0.1 0.6}

\newcommand{\Basestation}{%
  \pspolygon[fillstyle=solid, fillcolor=white](-2, -10)(2, -10)(0, -2)(-2, -10)
  \psline(0, -2)(0, 0)
  \psdots(0, 0)
  \rput{135}(3, 2.5){$\mathbf{\downarrow}$}
  \rput{225}(-2, 2.5){$\mathbf{\downarrow}$}
}
\newcommand{\Relaynode}{%
\rput(0, 0){\Huge{$\Ydown$}}%
}
\newcommand{\Hexagon}[1]{%
  \pspolygon[fillcolor=#1, fillstyle=solid](10, 0)(5, 8.6603)(-5, 8.6603)(-10, 0)(-5, -8.6603)(5, -8.6603)(10, 0)
}
\newcommand{\SiteA}[1]{%
  \rput(0, 0){\Hexagon{lightgray}}
  \rput(15, 8.6603){\Hexagon{lightblue}}
  \rput(0, 17.321){\Hexagon{lightred}}
  \rput(5, 14){\Basestation}

  \rput(0.1503, 5.8603){\Relaynode}
  \rput(5, 3.063){\Relaynode}
  \rput(9.8497, 5.8603){\Relaynode}
  \rput(9.8497, 11.4603){\Relaynode}
  \rput(5, 14.2603){\Relaynode}
  \rput(0.1503, 11.4603){\Relaynode}

  \rput(-4.2376, 3.327){\Relaynode}
  \rput(5, -2.0064){\Relaynode}
  \rput(14.2376, 3.3270){\Relaynode}
  \rput(14.2376, 13.9936){\Relaynode}
  \rput(5, 19.3270){\Relaynode}
  \rput(-4.2376, 13.9936){\Relaynode}
}
\newcommand{\SiteB}[1]{%
  \rput(0, 0){\Hexagon{lightblue}}
  \rput(15, 8.6603){\Hexagon{lightred}}
  \rput(15, -8.6603){\Hexagon{lightgray}}
  \rput(10, 6){\Basestation}

  \rput(5.1503, -2.1397){\Relaynode}
  \rput(10, -4.937){\Relaynode}
  \rput(14.8497, -2.1397){\Relaynode}
  \rput(14.8497, 3.4603){\Relaynode}
  \rput(10, 6.2603){\Relaynode}
  \rput(5.1503, 3.4603){\Relaynode}

  \rput(0.7624, -4.673){\Relaynode}
  \rput(10, -10.0064){\Relaynode}
  \rput(19.2376, -4.673){\Relaynode}
  \rput(19.2376, 5.9936){\Relaynode}
  \rput(10, 11.3270){\Relaynode}
  \rput(0.7624, 5.9936){\Relaynode}
}

\newcommand{\Bnull}{\mathbf{0}}
\newcommand{\Bone}{\mathbf{1}}

\newcommand{\Be}{\mathbf{e}}

\newcommand{\BG}{\mathbf{G}}

\newcommand{\Bh}{\mathbf{h}}

\newcommand{\BH}{\mathbf{H}}
\newcommand{\BI}{\mathbf{I}}

\newcommand{\BP}{\mathbf{P}}

\newcommand{\BS}{\mathbf{S}}

\newcommand{\BU}{\mathbf{U}}
\newcommand{\BV}{\mathbf{V}}
\newcommand{\BW}{\mathbf{W}}

\newcommand{\BE}{\mathbf{E}}

\newcommand{\Bp}{\mathbf{p}}

\newcommand{\Bn}{\mathbf{n}}

\newcommand{\Bv}{\mathbf{v}}
\newcommand{\BvBS}{\mathbf{v}^{\textnormal{BS}}}

\newcommand{\BvUE}{\mathbf{v}^{\textnormal{UE}}}

\newcommand{\Bs}{\mathbf{s}}

\newcommand{\Bx}{\mathbf{x}}

\newcommand{\By}{\mathbf{y}}

\newcommand{\inv}[1]{\left(#1\right)^{-1}}

\newcommand{\Pqq}{\mathbf{\Phi}_{\textnormal{qq}}}

\newcommand{\Pss}{\mathbf{\Phi}_{\textnormal{ss}}}

\newcommand{\LOG}{\textnormal{log}}

\newcommand{\MAX}{\textnormal{max}}

\newcommand{\DIAG}{\textnormal{diag}}

\newcommand{\MSE}{\textnormal{MSE}}
\newcommand{\TR}{\textnormal{tr}}

\newcommand{\dg}{\Delta}
\newcommand{\SPA}{\textnormal{ }}

\newcommand{\fD}{f_{\textnormal{D}}}
\newcommand{\fs}{f_{\textnormal{s}}}
\newcommand{\fc}{f_{\textnormal{c}}}

\newcommand{\Nb}{N_{\textnormal{b}}}

\newcommand{\NBS}{N_{\textnormal{BS}}}

\newcommand{\Ns}{N_{\textnormal{s}}}
\newcommand{\Nc}{N_{\textnormal{c}}}
\newcommand{\Nd}{N_{\textnormal{d}}}
\newcommand{\Nrank}{N_{\textnormal{rank}}}
\newcommand{\Nppos}{N_{\textnormal{ppos}}}

\newcommand{\SK}{\mathcal{K}}

\definecolor{areagray}{rgb}{0.95, 0.95, 0.95}
\definecolor{areadarkgray}{rgb}{0.8, 0.8, 0.8}

\definecolor{arealightgreen}{rgb}{0.9, 1, 0.9}
\definecolor{areagreen}{rgb}{0.7, 0.9, 0.7}
\definecolor{areadarkgreen}{rgb}{0.5, 0.8, 0.5}

\definecolor{arealightblue}{rgb}{0.9, 0.9, 1}
\definecolor{areablue}{rgb}{0.7, 0.7, 0.9}
\definecolor{areadarkblue}{rgb}{0.5, 0.5, 0.8}

\definecolor{arealightred}{rgb}{1, 0.9, 0.9}
\definecolor{areared}{rgb}{0.9, 0.7, 0.7}
\definecolor{areadarkred}{rgb}{0.8, 0.5, 0.5}

\definecolor{arealightpurple}{rgb}{1, 0.9, 1}
\definecolor{areapurple}{rgb}{0.9, 0.7, 0.9}
\definecolor{areadarkpurple}{rgb}{0.8, 0.5, 0.8}

\definecolor{arealightbrown}{rgb}{1, 0.95, 0.9}
\definecolor{areabrown}{rgb}{0.9, 0.8, 0.7}
\definecolor{areadarkbrown}{rgb}{0.8, 0.75, 0.5}

\definecolor{mypurple}{rgb}{0.6, 0, 0.6}
\definecolor{mybrown}{rgb}{1, 0.6, 0}
\definecolor{grey}{rgb}{0.5, 0.5, 0.5}

\acrodef{UL}{uplink}
\acrodef{DL}{downlink}
\acrodef{UT}{user terminal}
\acrodef{BS}{base station}
\acrodef{RRM}{radio resource management}
\acrodef{CSI}{channel state information}
\acrodef{PRB}{physical resource block}
\acrodef{TTI}{transmission time interval}
\acrodef{MAC}{multiple access channel}
\acrodef{BC}{broadcast channel}
\acrodef{TDM}{time division multiplex}

\title{Application Driven Joint Uplink-Downlink Optimization in Wireless Communications}
\author{
\authorblockN{Patrick Marsch, Peter Rost, and Gerhard Fettweis}
  \authorblockA{Technische Universität Dresden, Vodafone Chair Mobile Communications Systems,\\
  Email: \{marsch, rost, fettweis\}@ifn.et.tu-dresden.de}
  \thanks{The authors acknowledge the excellent cooperation of all project partners within the EASY-C project 
  and the support by the German Federal Ministry of Education and Research (BMBF).}%
}
\IEEEoverridecommandlockouts
\begin{document}
  \maketitle
  \begin{abstract}
    This paper introduces a new mathematical framework which is used to derive joint uplink/downlink achievable rate regions for multi-user spatial multiplexing between one base station and multiple terminals. The framework consists of two models: the first one is a simple transmission model for \ac{UL} and \ac{DL}, which is capable to give a lower bound on the capacity for the case that the transmission is subject to imperfect \ac{CSI}. A detailed model for concrete channel estimation and feedback schemes provides the parameter input to the former model and covers the most important aspects such as pilot design optimization, linear channel estimation, feedback delay, and feedback quantization.

    We apply this framework to determine optimal pilot densities and \ac{CSI} feedback quantity, given that a weighted sum of \ac{UL} and \ac{DL} throughput is to be maximized for a certain user velocity. We show that for low speed, and if \ac{DL} throughput is of particular importance, a significant portion of the \ac{UL} should be invested into CSI feedback. At higher velocity, however, \ac{DL} performance becomes mainly affected by \ac{CSI} feedback delay, and hence \ac{CSI} feedback brings little gain considering the inherent sacrifice of \ac{UL} capacity. We further show that for high velocities, it becomes beneficial to use no \ac{CSI} feedback at all, but apply random beamforming in the \ac{DL} and operate in time-division multiplex.
  \end{abstract}
  \section{Introduction}
    \subsection{Motivation}
      Mobile communication systems provide a diversity of high-quality mobile applications and services, which
      require a multitude of operation modes. Each mode is characterized among others by its requirements on latency, packet
      error rate, and supported rate. To satisfy these demands, currently deployed systems
      use different transport protocols, coding schemes, and modulation schemes. Furthermore,
      \ac{RRM} algorithms usually optimize the \ac{UL} and \ac{DL} data rate independently.

\medskip

      Consider a typical file download, which requires an \ac{UL}-\ac{DL} throughput ratio of $\nicefrac{R_\text{UL}}{R_\text{DL}}\ll 1$. In the context of multi-user MIMO, it is known that a strong DL requires CSI feedback from the \ac{UT} side to the \ac{BS} side, where precise multi-user precoding for spatial multiplexing can then be performed. By contrast, the upload of files and real-time video-streaming (for instance for remote surveillance) require a stronger \ac{UL} than \ac{DL}, which results in $\nicefrac{R_\text{UL}}{R_\text{DL}}\gg 1$.
      While the former examples represent asymmetric services, voice-over-IP or video-conferencing have symmetric rate demands,
      reflected by $\nicefrac{R_\text{UL}}{R_\text{DL}}=1$.
      
      \medskip
      
      In todays mobile communication systems, all three service classes use the same physical layer mode, although they have very contrary demands, and satisfy these by individual resource scheduling in \ac{UL} and \ac{DL}. In this work, we alternatively consider an application-driven multi-cross-layer approach, which \emph{jointly} optimizes both \ac{UL} and \ac{DL} not only on the upper layers but also on physical layer.

    \subsection{Outline of Main Contribution}
      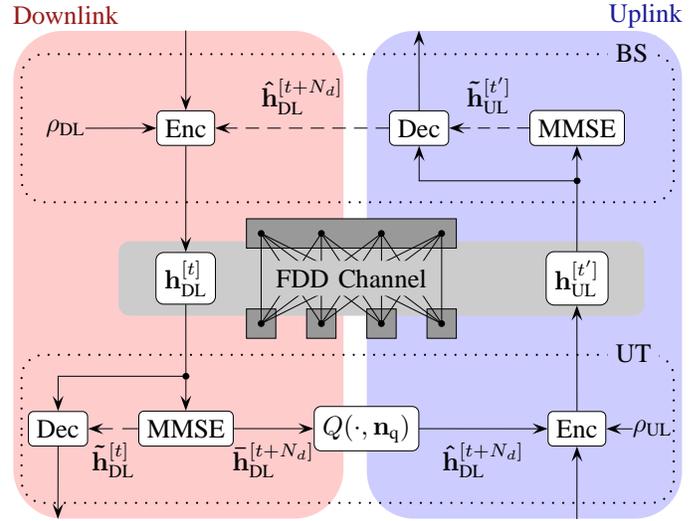
\begin{figure}[t]
	\centering
\begingroup
\unitlength=1mm
\begin{picture}(89, 68)(0, 0)
  \psset{xunit=1mm, yunit=1mm, linewidth=0.1mm}%
  \psset{arrowsize=3pt 4, arrowlength=1.4, arrowinset=.4}%
  \rput(9, 2){%
    \psframe[framearc=.3, fillcolor=verylightblue, fillstyle=solid, linecolor=white, linestyle=none](38, -2)(80, 63)%
    \psframe[framearc=.3, fillcolor=verylightred, fillstyle=solid, linecolor=white, linestyle=none](-9, -2)(35, 63)%
    \newrgbcolor{ChannelColor}{0.8 0.8 0.8}%
    \newrgbcolor{BoxesColor}{0.6 0.6 0.6}%
    \psframe[framearc=.3, fillcolor=ChannelColor, fillstyle=solid, linecolor=white, linestyle=none](5, 25)(75, 35)%
    \psframe[fillstyle=solid, fillcolor=BoxesColor](22, 34)(50, 38)%
    \psframe[fillstyle=solid, fillcolor=BoxesColor](22, 22)(26, 26)\psline(24, 24)(24, 36)\psline(24, 24)(32, 36)\psline(24, 24)(40, 36)\psline(24, 24)(48, 36)%
    \psframe[fillstyle=solid, fillcolor=BoxesColor](30, 22)(34, 26)\psline(32, 24)(24, 36)\psline(32, 24)(32, 36)\psline(32, 24)(40, 36)\psline(32, 24)(48, 36)%
    \psframe[fillstyle=solid, fillcolor=BoxesColor](38, 22)(42, 26)\psline(40, 24)(24, 36)\psline(40, 24)(32, 36)\psline(40, 24)(40, 36)\psline(40, 24)(48, 36)%
    \psframe[fillstyle=solid, fillcolor=BoxesColor](46, 22)(50, 26)\psline(48, 24)(24, 36)\psline(48, 24)(32, 36)\psline(48, 24)(40, 36)\psline(48, 24)(48, 36)%
    \psdots*[dotstyle=*](24, 36)(32, 36)(40, 36)(48, 36)
    \psdots*[dotstyle=*](24, 24)(32, 24)(40, 24)(48, 24)
    \rput[c](36, 30){\psframebox[fillcolor=ChannelColor, fillstyle=solid, linestyle=none]{FDD Channel}}%
    \rput[l](-9, 65){\textcolor{DLcolor}{Downlink}}%
    \rput[r](80, 65){\textcolor{ULcolor}{Uplink}}%
    \psframe[framearc=.3, fillstyle=none, linecolor=black, linewidth=0.3mm, linestyle=dotted](-8, 40)(79, 60)\rput[l](70, 60){\psframebox[fillcolor=verylightblue, fillstyle=solid, linestyle=none]{BS}}%
    \psframe[framearc=.3, fillstyle=none, linecolor=black, linewidth=0.3mm, linestyle=dotted](-8, 0)(79, 20)\rput[l](70, 20){\psframebox[fillcolor=verylightblue, fillstyle=solid, linestyle=none]{UT}}%
    %
    \rput(-3, -2){\rnode{SourceMessageW}{~}}%
    \rput(66, -2){\rnode{SourceMessageU}{~}}%
    \rput(-3, 10){\rnode{SourceDecoder}{\psframebox[fillcolor=white, fillstyle=solid, framearc=.3]{Dec}}}%
    \rput(14, 10){\rnode{SourceEstimator}{\psframebox[fillcolor=white, fillstyle=solid, framearc=.3]{MMSE}}}%
    \rput(38, 10){\rnode{SourceQuantizer}{\psframebox[fillcolor=white, fillstyle=solid, framearc=.3]{$Q(\cdot, \Bn_{\textnormal{q}})$}}}%
    \rput(66, 10){\rnode{SourceEncoder}{\psframebox[fillcolor=white, fillstyle=solid, framearc=.3]{Enc}}}%
    \rput(76, 10){\rnode{PilotsUL}{$\rho_{\textnormal{UL}}$}}%
    %
    \ncline[linestyle=dashed]{->}{SourceEstimator}{SourceDecoder}\naput{$\MatrixTime{\tilde h}{t}_{\text{DL}}$}%
    \ncline{->}{SourceEstimator}{SourceQuantizer}\nbput{$\MatrixTime{\bar h}{t+N_d}_{\text{DL}}$}%
    \ncline{->}{SourceQuantizer}{SourceEncoder}\nbput{$\MatrixTime{\hat{h}}{t+N_d}_{\text{DL}}$}%
    \ncline{->}{PilotsUL}{SourceEncoder}%
    \ncline{->}{SourceDecoder}{SourceMessageW}%
    \ncline{->}{SourceMessageU}{SourceEncoder}%
    %
    \dotnode(14, 17){ChannelOutput1}%
    \rput(14, 30){\rnode{ChannelDL}{\psframebox[fillcolor=white, fillstyle=solid, framearc=.3]{$\MatrixTime{h}{t}_{\text{DL}}$}}}%
    \rput(66, 30){\rnode{ChannelUL}{\psframebox[fillcolor=white, fillstyle=solid, framearc=.3]{$\MatrixTime{h}{t'}_{\text{UL}}$}}}%
    \ncline{-}{ChannelDL}{ChannelOutput1}
    \ncangle[angleB=90]{->}{ChannelOutput1}{SourceDecoder}%
    \ncline{->}{ChannelOutput1}{SourceEstimator}%
    \ncline{->}{SourceEncoder}{ChannelUL}
    %
    \rput(66, 50){\rnode{BSEstimator}{\psframebox[fillcolor=white, fillstyle=solid, framearc=.3]{MMSE}}}%
    \rput(45, 50){\rnode{BSDecoder}{\psframebox[fillcolor=white, fillstyle=solid, framearc=.3]{Dec}}}%
    \rput(14, 50){\rnode{BSEncoder}{\psframebox[fillcolor=white, fillstyle=solid, framearc=.3]{Enc}}}%
    \rput(-2, 50){\rnode{PilotsDL}{$\rho_{\textnormal{DL}}$}}%
    \rput(14, 63){\rnode{BSMessageW}{~}}%
    \rput(45, 63){\rnode{BSMessageU}{~}}%
    \dotnode(66, 43){ChannelOutputBS}%
    %
    \ncline{->}{BSEncoder}{ChannelDL}
    \ncline{->}{PilotsDL}{BSEncoder}%
    \ncline{->}{BSMessageW}{BSEncoder}%
    \ncline[linestyle=dashed]{->}{BSDecoder}{BSEncoder}\nbput{$\MatrixTime{\hat{h}}{t+N_d}_{\text{DL}}$}%
    \ncline[linestyle=dashed]{->}{BSEstimator}{BSDecoder}\nbput{$\MatrixTime{\tilde h}{t'}_{\text{UL}}$}%
    \ncline{->}{BSDecoder}{BSMessageU}%
    \ncline{-}{ChannelUL}{ChannelOutputBS}
    \ncline{->}{ChannelOutputBS}{BSEstimator}%
    \ncangle[angleB=-90]{->}{ChannelOutputBS}{BSDecoder}%

  }
\end{picture}
\endgroup
	\caption{Information flow between uplink and downlink.}
	\label{fig:system.model:information.flow}
      \end{figure}
      This paper presents 
      \begin{enumerate}
	\item
	  a simplified model of an \ac{UL} and \ac{DL} transmission over a frequency-flat channel for capacity calculation 
	  under CSI imperfectness at \ac{BS} and \ac{UT} side, 
	\item
	  a detailed model that yields the extent of \ac{CSI} imperfectness for a concrete OFDM-based channel estimation and \ac{CSI} feedback scheme, applied to a channel with a certain dispersiveness in time and frequency, and
	\item 
	  an \ac{UL}/\ac{DL} tradeoff discussion, which analyzes the best choice of pilot density and \ac{CSI} feedback amount, given that a weighted sum-rate of \ac{UL} and \ac{DL} is to be maximized.
      \end{enumerate}
      
      In general, we consider the communication between a \ac{BS} with $\NBS$ receive and transmit antennas, and $K$ \ac{UT}s with one receive and transmit antenna each, as depicted in Fig.~\ref{fig:system.model:information.flow}. The set of \ac{UT}s is defined by $\SK = \{1..K\}$. Within this setup, we are particularly interested in optimizing the following parameters:
      \begin{itemize}
	\item the pilot density $\rho_{\textnormal{UL}} \in \DefSet{R}^+$ in the \ac{UL},
	\item the pilot density $\rho_{\textnormal{DL}} \in \DefSet{R}^+$
	  in the \ac{DL}, and
	\item the amount of \ac{CSI} feedback $N_b \in \DefSet{R}^+_0$ in bits per channel coefficient and \ac{PRB}.
      \end{itemize}
      We generally assume a scenario-dependent optimized layout of \ac{UL} and \ac{DL} pilots in time and frequency, so that the pilot densities $\rho_{\textnormal{UL}}$ and $\rho_{\textnormal{DL}}$ are sufficient as optimization parameters. Fig. \ref{fig:system.model:information.flow} shows the information flow in the considered system. The right side illustrates the \ac{UL} and the left side the \ac{DL}. Via MMSE channel estimation, each \ac{UT} generates both a channel estimate $\MatrixTime{\tilde{h}}{t}_{\text{DL}}$ for data detection as well as a channel prediction $\MatrixTime{\bar{h}}{t+\Nd}_{\text{DL}}$ for \ac{CSI} feedback. The latter is quantized to $\MatrixTime{\hat{h}}{t+\Nd}_{\text{DL}}$, introducing quantization noise $\Matrix{n}_q$, before it is sent by the \ac{UL} to the \ac{BS} (introducing a delay of $\Nd$ \ac{TTI}). At the \ac{BS} side, an MMSE estimator yields the channel estimate $\MatrixTime{\tilde{h}}{t}_{\text{UL}}$, which is used to decode the \ac{UL} transmission. This includes the \ac{CSI} feedback from the \ac{UT} side, which is then used to perform linear precoding to improve the \ac{DL} throughput. 

    \subsection{Previous Work}  
      To the best of the authors' knowledge, this is the first work considering a model which incorporates all major aspects of imperfect
      channel state information in a bi-directional, multi-user wireless communication system. The problem of imperfect \ac{CSI} at the transmitter as a function of the \ac{CSI} at the receiver has been first considered by Caire and Shamai in \cite{Caire.Shamai.TransIT.1999}. In \cite{Santipach.Honig.ISIT.2006}, Santipach and Honig considered
      both imperfect channel estimation and quantized channel feedback. More recently, Kobayashi \etal~\cite{Kobayashi.Caire.Jindal.ISIT.2008} analyze
      a system where the \ac{DL} throughput depends on the channel estimation at the receiver and the amount of feedback, which is constrained
      by a given \ac{UL} capacity (independent of the channel estimation at the \ac{BS}). While \cite{Kobayashi.Caire.Jindal.ISIT.2008}
      assumes a channel static during both \ac{UL} and \ac{DL}, the same authors consider FDD models in \cite{Kobayashi.Jindal.Caire.arxiv.2009},
      which are also in the focus of this work. More specifically, \cite{Kobayashi.Jindal.Caire.arxiv.2009} analyzes the performance of digital and analog feedback, evaluates the jointly achievable \ac{UL}/\ac{DL} rate region, and considers user scheduling in addition to a feedback optimization. 

    \subsection{Outline of Paper}
      In Section \ref{sec:simp_model}, we introduce a simplified transmission model for \ac{UL} and \ac{DL} which yields lower bounds on the capacity for transmission under imperfect \ac{CSI}. A detailed model for channel estimation and \ac{CSI} feedback will be introduced in Section \ref{sec:detailed_mod} and provides the parameter values for the simpler model in Section \ref{sec:simp_model}. We use both models in Section \ref{sec:joint.rate.region} to discuss the joint optimization of \ac{UL} and \ac{DL} throughput and in Section \ref{sec:results} to evaluate the individual parameter values, which achieve optimal sum-rate under a given \ac{UL}/\ac{DL} throughput ratio. The paper is concluded in Section \ref{sec:conclusions}.

\section{Simplified Model for Capacity Calculation}\label{sec:simp_model}
In our simplified model for capacity calculation, we assume that all entities are perfectly synchronized in time and frequency and that transmission takes place over a frequency-flat channel. We generally assume that all involved signals are realizations of Gaussian processes.   

\subsection{Uplink}

In the \ac{UL}, we are facing a \ac{MAC}, and model the transmission of each symbol as
\begin{equation}
\label{e:TRANSMISSION_UL}
\By = \BH^{\textnormal{UL}} \Bx + \Bn,
\end{equation}

\noindent where $\By \in \mathbb{C}^{[\NBS \times 1]}$ are the signals received at the BS antennas, $\BH^{\textnormal{UL}} \in \mathbb{C}^{[\NBS \times K]}$ is the channel matrix, $\Bx \in \mathbb{C}^{[K \times 1]}$ are the signals transmitted from the $K$ terminals with $\BP = E\{\Bx\Bx^H\}$, and $\Bn \sim \mathcal{N}_{\mathbb{C}}(\Bnull, \sigma_{\textnormal{UL}}^2\BI)$ is receiver-side noise. The \ac{UL} is subject to a per-UT power constraint which we state as $\forall \SPA k \in \mathcal{K} \SPA : \SPA E\{x_k x_k^H\} \leq P^{\textnormal{max}}_{\textnormal{UL}}$. Let the channel knowledge at the BS side be
\begin{equation}
\label{e:CHNEST_ERROR_UL}
\tilde{\BH}^{\textnormal{UL}} = \BH^{\textnormal{UL}} + \BE^{\textnormal{UL}},
\end{equation} 

\noindent which corresponds to the actual channel {\em plus} a channel estimation error $\BE^{\textnormal{UL}} \in \mathbb{C}^{[\NBS \times K]}$. We further assume that all entries in $\BE^{\textnormal{UL}}$ are uncorrelated Gaussian variables with $E\{vec(\BE^{\textnormal{UL}})vec(\BE^{\textnormal{UL}})^H\} = \sigma_{\textnormal{UL,BS}}^2$. The latter variance can be obtained through the Kramer-Rao lower bound~\cite{Kay_BOOK93} as $\sigma_{\textnormal{UL,BS}}^2 = \sigma^2_{\textnormal{UL}}/(N_{\textnormal{pilots}}\cdot p_{\textnormal{pilots}})$, if channel estimation has been performed based on the transmission of $N_{\textnormal{pilots}}$ pilots of power $p_{\textnormal{pilots}}$. It has been shown in~\cite{Marsch_ICC09} that we can find an inner bound on the capacity region connected to the transmission in~\eqref{e:TRANSMISSION_UL} by observing a modified transmission
\begin{equation}
\label{e:TRANSMISSION_UL_MODIFIED}
\By = \BH^{\textnormal{UL,eff}} \Bx + \Bv + \Bn,
\end{equation}

\noindent where the channel is reduced in power to an {\em effective} channel
\begin{equation}
\label{e:UL_EFFECTIVE_CHANNEL}
\forall \SPA i,j : \SPA h^{\textnormal{UL,eff}}_{i,j} = \frac {h^{\textnormal{UL}}_{i,j}} {\sqrt{1 + \sigma_{\textnormal{UL,BS}}^2 \left/ E\left\{\left|h^{\textnormal{UL}}_{i,j}\right|^2\right\}\right.}},
\end{equation}

\noindent and with an additional channel estimation related noise term
\begin{multline}
\label{e:UL_EFFECTIVE_NOISE}
E\left\{\Bv \Bv^H \right\} = \DIAG \left( \DIAG \left( \bar{\BE}^{\textnormal{UL,eff}} \BP \left(\bar{\BE}^{\textnormal{UL,eff}}\right)^H \right) \right),
\\ \textnormal{ where } \SPA \forall \SPA i,j : \SPA \bar{e}^{\textnormal{UL,eff}}_{i,j} = \sqrt{\frac {E\left\{\left|h^{\textnormal{UL}}_{i,j}\right|^2\right\} \cdot \sigma_{\textnormal{UL,BS}}^2}{E\left\{\left|h^{\textnormal{UL}}_{i,j}\right|^2\right\} + \sigma_{\textnormal{UL,BS}}^2}}.
\end{multline}

Briefly, the derivation of~\eqref{e:UL_EFFECTIVE_CHANNEL} and~\eqref{e:UL_EFFECTIVE_NOISE} is based on the fact that the channel estimation noise is treated as a Gaussian variable, leading to an overestimation of its detrimental impact~\cite{Medard_IEEETRANS00}.

\medskip

The sum-rate of all UTs can now be lower-bounded as~\cite{Marsch_ICC09}
\begin{eqnarray}
\label{e:SUMRATE_UL}
R_{\textnormal{UL}} &\!\!\leq\!\!& \!\!\!\!\underset{\BP - P_{\textnormal{UL}}^{\textnormal{max}}\BI \preceq 0}{\MAX} \LOG_2 \left| \BI \! + \! \mathbf{\Phi}^{-1} \BH^{\textnormal{UL,eff}} \BP \left( \BH^{\textnormal{UL,eff}} \right)^H \right| \\
\label{e:SUMRATE_UL2}
\textnormal{with} \SPA \mathbf{\Phi} &\!\!=\!\!& \sigma_{\textnormal{UL}}^2 \BI + \DIAG \left( \DIAG \left(\bar{\BE}^{\textnormal{UL,eff}} \BP \left(\bar{\BE}^{\textnormal{UL,eff}} \right)^H \right) \right).
\end{eqnarray}

Note that~\eqref{e:SUMRATE_UL} requires optimization over all power allocations $\BP$ that fulfill the individual power constraints. Under perfect CSIR, the sum-rate maximizing strategy is to let all UTs transmit at maximum power, which, however, is not necessarily the case under imperfect CSIR, as the channel estimation related noise term in~\eqref{e:SUMRATE_UL2} depends on $\BP$. 

\subsection{Downlink}

The DL corresponds to a \ac{BC}, where the transmission of each symbol can be stated as
\begin{equation}
\label{e:TRANSMISSION_DL}
\By = \left(\BH^{\textnormal{DL}}\right)^H \Bs = \left(\BH^{\textnormal{DL}}\right)^H \BW \Bx + \Bn,
\end{equation}

\noindent where $\By \in \mathbb{C}^{[K \times 1]}$ are now the signals received by the UTs, $\BH^{\textnormal{DL}} \in \mathbb{C}^{[\NBS \times K]}$ is the \ac{DL} channel matrix, $\BW \in \mathbb{C}^{[\NBS \times K]}$ is a precoding matrix, $\Bx \in \mathbb{C}^{[K \times 1]}$ are signals to be transmitted to the $K$ UTs, and $\Bn \sim \mathcal{N}_{\mathbb{C}}(\Bnull, \sigma_{\textnormal{DL}}^2 \BI)$ is UT-side noise. $\BH^{\textnormal{DL}}$ is usually different from the \ac{UL} channel $\BH^{\textnormal{UL}}$ due to different frequencies and hence different path loss and scattering. We consider a sum-power constraint $\TR\{E\{\Bs\Bs^H\}\} \leq P^{\textnormal{tot}}_{\textnormal{DL}}$, and assume that the UTs have the (distributed) channel estimate
\begin{equation}
\label{e:CHNEST_ERROR_DL_UE}
\tilde{\BH}^{\textnormal{DL}} = \BH^{\textnormal{DL}} + \BE^{\textnormal{DL,UT}}, 
\end{equation}

\noindent with $E\{vec(\BE^{\textnormal{DL,UT}})vec(\BE^{\textnormal{DL,UT}})^H\} = \sigma_{\textnormal{DL,UT}}^2$, and that the BS side has an even noisier channel estimate
\begin{equation}
\label{e:CHNEST_ERROR_DL_BS}
\hat{\BH}^{\textnormal{DL}} = \sqrt{\alpha} \left(\BH^{\textnormal{DL}} + \BE^{\textnormal{DL,UT}}\right) + \BE^{\textnormal{DL,BS}},
\end{equation}

\noindent with $E\{vec(\BE^{\textnormal{DL,BS}})vec(\BE^{\textnormal{DL,BS}})^H\} = \sigma_{\textnormal{DL,BS}}^2$. Scaling factor $\alpha$ ensures that the power of the channel estimate $\hat{\BH}^{\textnormal{DL}}$ at the BS side corresponds to that of $\tilde{\BH}^{\textnormal{DL}}$ again~\cite{CoverThomas_BOOK06}. The model is motivated through the fact that in an FDD system, CSI at the transmitter side can only be obtained through feedback from the receiver side. Hence, it is always strictly less accurate (due to quantization and delay) than at the receiver side. Again it is possible to model the impact of imperfect \ac{CSI} through the observation of a modified transmission equation~\cite{Marsch_GLOBECOM09}
\begin{multline}
\By = \overbrace{\BH^{\textnormal{DL,eff}} \BW \Bx}^{\textnormal{Controllable term}} + \BvUE + \BvBS + \Bn \SPA\\
\textnormal{with} \SPA \forall \SPA i,j : \SPA h^{\textnormal{DL,eff}}_{i,j} = h^{\textnormal{DL}}_{i,j} \cdot \sqrt{\frac {E\left\{\left|h^{\textnormal{DL}}_{i,j}\right|^2\right\} - \sigma_{\textnormal{DL,BS}}^2} {E\left\{\left|h^{\textnormal{DL}}_{i,j}\right|^2\right\} + \sigma_{\textnormal{DL,UT}}^2}}, \\
\BvUE \sim \mathcal{N}_{\mathbb{C}}\left(\Bnull, \dg\left(\bar{\BE}^{\textnormal{DL,UT}} \Pss \left(\bar{\BE}^{\textnormal{DL,UT}}\right)^H \right) \right),\\
\BvBS \sim \mathcal{N}_{\mathbb{C}}\left(\Bnull, \dg\left(\bar{\BE}^{\textnormal{DL,BS}} \Pss \left(\bar{\BE}^{\textnormal{DL,BS}}\right)^H \right) \right),\\
\forall \SPA i,j : \SPA \bar{e}^{\textnormal{DL,BS}}_{i,j} = \sqrt{ \frac{ \sigma_{\textnormal{DL,BS}}^2 \left(E\left\{|h^{\textnormal{DL}}_{i,j}|^2\right\}\right)^2}{E\left\{|h^{\textnormal{DL}}_{i,j}|^2\right\} + \sigma_{\textnormal{DL,UT}}^2}} \\ \textnormal{and} \SPA \forall \SPA i,j : \SPA \bar{e}^{\textnormal{DL,UT}}_{i,j} = \sqrt{\frac {E\left\{|h^{\textnormal{DL}}_{i,j}|^2\right\} \sigma_{\textnormal{DL,UT}}^2} {E\left\{|h^{\textnormal{DL}}_{i,j}|^2\right\} + \sigma_{\textnormal{DL,UT}}^2}}  \label{e:TRANSMISSION_MODIFIED_DL}
\end{multline}

Here, $\BvUE$ is a noise term related to imperfect CSI at BS and \ac{UT} side, while $\BvBS$ is connected to additional CSI imperfectness at the BS side~\cite{Marsch_GLOBECOM09}. Note that, as in the \ac{UL}, the modified transmission equation in~\eqref{e:TRANSMISSION_MODIFIED_DL} implies that statistical knowledge on the channel and estimation error is given at both BS and \ac{UT} side. While direct capacity region calculation in a \ac{BC} is tedious,\ac{UL}/\ac{DL} duality~\cite{Vishwanath.Jindal.Goldsmith.TransIT.2003} can strongly facilitate this, and is also applicable in the context of imperfect CSI~\cite{Marsch_GLOBECOM09}. We are then facing a dual \ac{UL} transmission under a sum power constraint, where the sum rate is given as in~\eqref{e:SUMRATE_DL}, optimized over the dual \ac{UL} power $\tilde{\Bp} \in \mathbb{R}_0^{+[K \times 1]}$ s.t. $\tilde{\Bp}^T\Bone \leq P_{\textnormal{DL}}^{\textnormal{tot}}$. In the denominator of the large fraction in~\eqref{e:SUMRATE_DL}, the first term is due to inter-UT interference, the second due to imperfect CSI at receiver and transmitter side, and the third due to additional CSI imperfectness at the \ac{BS} side.
\begin{figure*}
\begin{equation}
\label{e:SUMRATE_DL}
R_{\textnormal{DL}} \leq \underset{\tilde{\BP}}{\MAX}  \sum\limits_{k=1}^{K} \LOG_2 \left| \BI + \frac {\tilde{p}_k \Bh^{\textnormal{DL,eff}}_k \left(\Bh^{\textnormal{DL,eff}}_k\right)^H} {\sum\limits_{j \neq k} \tilde{p}_j \Bh^{\textnormal{DL,eff}}_j \left(\Bh^{\textnormal{DL,eff}}_j\right)^H \!\!\!\!\!\! + \!\! \sum\limits_{j=1}^{K} \tilde{p}_j \dg\left(\bar{\Be}^{\textnormal{DL,UT}}_j \left(\bar{\Be}^{\textnormal{DL,UT}}_j\right)^H \right) \!\! + \!\! \sum\limits_{j \neq k} \tilde{p}_j \dg\left(\bar{\Be}^{\textnormal{DL,BS}}_j \left(\bar{\Be}^{\textnormal{DL,BS}}_j\right) \right) \!\! + \! \sigma_{\textnormal{DL}}^2 \BI} \right|
\end{equation}
\end{figure*}

\subsection{TDM as an Alternative in the DL}
In~\eqref{e:TRANSMISSION_MODIFIED_DL}, the power of the effective channel goes to zero as the \ac{CSI} at the transmitter side diminishes. However, our model does not capture the fact that the system can always operate in \ac{TDM} and perform random beamforming to each \ac{UT} successively. The average sum-rate achievable with TDM is given as
\begin{equation}
\label{e:SUMRATE_DL_TDM}
R_{\textnormal{DL,TDM}} \leq \frac {1}{K} \sum\limits_{k=1}^K \LOG_2 \left( 1 + \frac{\left| h^{\textnormal{DL,eff,TDM}}_k \right|^2}{\sigma_{\textnormal{TDM},k}^2 + \sigma_{\textnormal{DL}}^2} \right) 
\end{equation}

\noindent where $h^{\textnormal{DL,eff,TDM}}_k$ is again a power-reduced effective channel, and $\sigma_{\textnormal{TDM},k}^2$ is a noise term connected to imperfect receiver-side channel knowledge at UT $k$, given as $\forall \SPA k \in \SK$ :
\begin{eqnarray}
\label{e:SUMRATE_DL_TDM2}
h^{\textnormal{DL,eff,TDM}}_k &\!\!=\!\!& \sqrt{\frac{P_{\textnormal{DL}}^{\textnormal{tot}}}{\NBS}} \Bh^{\textnormal{DL}}_k \Bone \sqrt{\frac{\frac{P_{\textnormal{DL}}^{\textnormal{tot}}}{\NBS} E\left\{ \left(\Bh^{\textnormal{DL}}_k\right)^H \!\!\Bh^{\textnormal{DL}}_k \right\}}{\frac{P_{\textnormal{DL}}^{\textnormal{tot}}}{\NBS} E\left\{ \left(\Bh^{\textnormal{DL}}_k\right)^H \!\!\Bh^{\textnormal{DL}}_k \right\} + \sigma_{\textnormal{DL,UT}}^2}} \nonumber \\
\sigma_{\textnormal{TDM},k}^2 &\!\!=\!\!& \frac{\sigma_{\textnormal{DL,UT}}^2} {\frac{P_{\textnormal{DL}}^{\textnormal{tot}}}{\NBS} E\left\{ \left(\Bh^{\textnormal{DL}}_k\right)^H \!\!\Bh^{\textnormal{DL}}_k \right\} + \sigma_{\textnormal{DL,UT}}^2}.  
\end{eqnarray} 

\medskip

A special aspect of TDM is that only one channel coefficient has to be estimated by each UT, namely the coefficient connected to the {\em effective} channel after random precoding at the BS side, reducing the pilot overhead in the DL. In the remainder of this work, we will always consider both instantaneous spatial multiplexing in the DL as well as TDM, and choose the better of both for any given scenario. Clearly, the value of sum-rate terms obtained through~\eqref{e:SUMRATE_UL},~\eqref{e:SUMRATE_DL} and~\eqref{e:SUMRATE_DL_TDM} depends strongly on the choice of terms $\sigma_{\textnormal{UL,BS}}^2$, $\sigma_{\textnormal{DL,UT}}^2$ and $\sigma_{\textnormal{DL,BS}}^2$, which in a practical system depend on the exact channel estimation and CSI feedback scheme as well as on the terminal speed $v$ and maximum delay spread $\tau_{\textnormal{max}}$. We hence require a lookup table providing
\begin{equation}
\label{e:LOOKUP}
\left[ \sigma_{\textnormal{UL,BS}}^2, \sigma_{\textnormal{DL,UT}}^2, \sigma_{\textnormal{DL,BS}}^2 \right] = f\left( \rho_{UL}, \rho_{DL}, \Nb, v, \tau_{\textnormal{max}} \right).
\end{equation}

\section{Detailed Model for Channel Estimation and CSI Feedback}\label{sec:detailed_mod}
           \begin{figure}[t]
	\begingroup
\unitlength=1mm
\begin{picture}(98, 35)(0, 0)

  \psset{xunit=0.8mm, yunit=0.8mm, linewidth=0.2mm}
    
  \rput(0, -10){%
    
    \rput(30, 55){\rnode{pilots_csifb}{$\BS$}} 
    \rput(50, 55){\rnode{chnestnoise_csifb}{$\Bn$}} 
    \rput(85, 55){\rnode{mmsefilter_csifb}{$\BG$ or $\BG_{\textnormal{P}}$}} 
      
    \rput(10, 45){\rnode{actchannel}{$\mathbf{h}^{[t]}$}} 
    \rput(30, 45){\rnode{mult_pilots_csifb}{\scalebox{2}{$\otimes$}}} 
    \rput(50, 45){\rnode{plus_chnestnoise_csifb}{\scalebox{2}{$\oplus$}}} 
    \rput(85, 45){\rnode{mult_mmsefilter_csifb}{\scalebox{2}{$\otimes$}}} 
    \rput(100, 45){\rnode{estchannel_ue}{$\tilde{\mathbf{h}}^{[t]}$}} 
    
    \rput(10, 38){\scalebox{0.7}{Actual channel}} 
    \rput(30, 38){\scalebox{0.7}{Pilot pattern}} 
    \rput(50, 38){\scalebox{0.7}{Chn. est. noise}} 
    
    \rput(30, 30){\rnode{rankrec}{$\BV$}} 
    \rput(70, 30){\rnode{quantnoise}{$\Bn_{\textnormal{q}}$}} 
    \rput(100, 20){\rnode{rankred}{$\BV^H$}} 
      
    \rput(10, 20){\rnode{estchannel_bs}{$\hat{\mathbf{h}}^{[t]}$}} 
    \rput(30, 20){\rnode{mult_rankrec}{\scalebox{2}{$\otimes$}}} 
    \rput(50, 20){\rnode{feedback}{\psframebox[fillstyle=solid, fillcolor=white]{$z^{-N_d}$}}}
    \rput(70, 20){\rnode{plus_quantnoise}{\scalebox{2}{$\oplus$}}} 
    \rput(85, 20){\rnode{mult_rankred}{\scalebox{2}{$\otimes$}}} 
    
    \rput(10, 13){\scalebox{0.7}{CSIT}} 
    \rput(30, 13){\scalebox{0.7}{Rank}} 
    \rput(30, 10){\scalebox{0.7}{reconstr.}} 
    \rput(70, 13){\scalebox{0.7}{Quant.}} 
    \rput(70, 10){\scalebox{0.7}{noise}} 
    \rput(85, 13){\scalebox{0.7}{Rank red.}} 
    \rput(74, 42){\scalebox{0.7}{MMSE filter}} 
    }
    
  \ncline[linestyle=solid, nodesep=3pt]{->}{effchannel}{mult_pilots_dec}
  \ncline[linestyle=solid, nodesep=3pt]{->}{pilots_dec}{mult_pilots_dec}
  \ncline[linestyle=solid, nodesep=3pt]{->}{mult_pilots_dec}{plus_chnestnoise_dec}
  \ncline[linestyle=solid, nodesep=3pt]{->}{chnestnoise_dec}{plus_chnestnoise_dec}
  \ncline[linestyle=solid, nodesep=3pt]{->}{actchannel}{mult_pilots_csifb}
  \ncline[linestyle=solid, nodesep=3pt]{->}{pilots_csifb}{mult_pilots_csifb}
  \ncline[linestyle=solid, nodesep=3pt]{->}{mult_pilots_csifb}{plus_chnestnoise_csifb}
  \ncline[linestyle=solid, nodesep=3pt]{->}{chnestnoise_csifb}{plus_chnestnoise_csifb}
  
  \ncline[linestyle=solid, nodesep=3pt]{->}{plus_chnestnoise_dec}{mult_mmsefilter_dec}
  \ncline[linestyle=solid, nodesep=3pt]{->}{mmsefilter_dec}{estchannel_ue}
  \ncline[linestyle=solid, nodesep=3pt]{->}{mmsefilter_dec}{mult_mmsefilter_dec}
  \ncline[linestyle=solid, nodesep=3pt]{->}{mult_mmsefilter_csifb}{estchannel_ue}
  \ncline[linestyle=solid, nodesep=3pt]{->}{plus_chnestnoise_csifb}{mult_mmsefilter_csifb}
  \ncline[linestyle=solid, nodesep=3pt]{->}{mmsefilter_csifb}{mult_mmsefilter_csifb}
  \ncline[linestyle=solid, nodesep=3pt]{->}{mult_mmsefilter_csifb}{mult_rankred}\Aput*{$\bar{\Bh}^{[t+\Nd]}$}
  \ncline[linestyle=solid, nodesep=3pt]{->}{rankred}{mult_rankred}
  \ncline[linestyle=solid, nodesep=3pt]{->}{mult_rankred}{plus_quantnoise}
  \ncline[linestyle=solid, nodesep=3pt]{->}{quantnoise}{plus_quantnoise}
  \ncline[linestyle=solid, nodesep=3pt]{->}{plus_quantnoise}{feedback}
  \ncline[linestyle=solid, nodesep=3pt]{->}{feedback}{mult_rankrec}
  \ncline[linestyle=solid, nodesep=3pt]{->}{rankrec}{mult_rankrec}
  \ncline[linestyle=solid, nodesep=3pt]{->}{mult_rankrec}{estchannel_bs}
      
  	      
  \end{picture}
\endgroup
	\caption{Detailed channel estimation and CSI feedback model.}
	\label{fig:csifeedback}
      \end{figure}
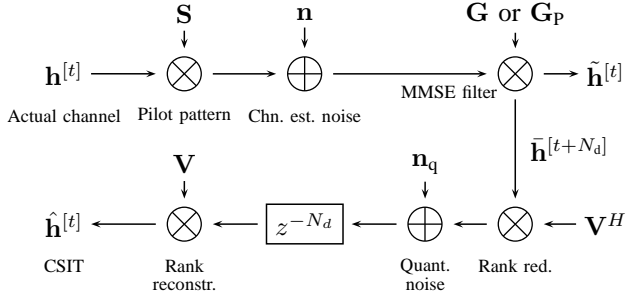
To obtain~\eqref{e:LOOKUP}, let us consider a particular OFDMA system as it is used in the DL of LTE Release 8~\cite{McCoy_2007}, with a symbol rate $\fs = 14$ kHz and a sub-carrier spacing $\Delta F = 15$ kHz. For simplicity, we assume that OFDMA is also used in the \ac{UL} (rather than SC-FDMA). Channel estimation is performed individually for each \ac{PRB} spanning $\Ns = 14$ OFDM symbols times $\Nc=12$ sub-carriers, hence $168$ channel accesses. The channel estimation performance could be improved using a larger observation window, which raises complexity issues. \ac{CSI} feedback is also performed on a \ac{PRB} basis, but with the option of using successive schemes that exploit the channel correlation over multiple TTIs.

\medskip

See the detailed channel estimation and CSI feedback model in Fig.~\ref{fig:csifeedback}. Different from before, we now consider the vector $\Bh^{[t]} \in \mathbb{C}^{[\Ns\Nc \times 1]}$, which stacks all channel realizations connected to the link between one BS antenna $a$ and one UT $k$ for all channel accesses within a \ac{PRB} at time $t$ into one vector. As we assume all links to be uncorrelated, channel estimation and CSI feedback have to be performed individually for each channel coefficient, where we omit the indices $a$ and $k$ in our notation for brevity. Matrix $\BS \in \{0,1\}^{[\Ns\Nc \times \Nppos]}$ indicates the $\Nppos$ pilot positions within the \ac{PRB}. Channel estimation is assumed to be subject to uncorrelated Gaussian noise $\Bn \sim \mathcal{N}_{\mathbb{C}}(\Bnull, \sigma_{\textnormal{p}}^2 \BI)$.

\subsection{Channel Estimation}
At the receiver side, a channel estimate $\tilde{\Bh}^{[t]} \in \mathbb{C}^{[\Ns\Nc \times 1]}$ for each OFDM symbol in a \ac{PRB} at time $t$ is obtained by applying an MMSE filter $\BG$ to the received pilot symbols:
\begin{equation}
\label{e:CSIR_APPLYFILTER}
\tilde{\Bh}^{[t]} = \BG \left( \BS \Bh^{[t]} + \Bn \right) 
\end{equation}
where the filter matrix is given as~\cite{Hoeher_BOOK97, Hoeher_ICASSP97}
\begin{equation}
\label{e:CSIR_MMSEFILTER}
\BG = \mathbf{\Phi}_{\textnormal{hh}} \left( 0 \right) \BS^H \inv{ \BS \mathbf{\Phi}_{\textnormal{hh}} \left( 0 \right) \BS^H + \sigma_{\textnormal{p}}^2 \BI}.
\end{equation}
Under the assumption of a wide-sense stationary uncorrelated scattering (WSSUS) channel fading model~\cite{DentCroft_LETTERS93},
the filter $\BG$ exploits the correlation of $\Bh$ in time and frequency, and is given as~\cite{Hoeher_BOOK97,Hoeher_ICASSP97}
\begin{eqnarray}
\mathbf{\Phi}_{\textnormal{hh}} \left( \Nd \right) &\!\!\!=\!\!\!& E\left\{\Bh^{[t-\Nd]} \left(\Bh^{[t]}\right)^H \right\} \\
&\!\!\!=\!\!\!& E\left\{ \left| h \right|^2 \right\} \cdot \left( \mathbf{\Pi}_{\textnormal{T}} \left( \Nd \right) \otimes \mathbf{\Pi}_{\textnormal{F}} \right),
\end{eqnarray}
\begin{figure*}
\begin{equation}
\label{e:PI_T}
\mathbf{\Pi}_{\textnormal{T}} \left( \Nd \right) = \left[ \begin{array}{cccc} \!\! J_0 \! \left(2 \pi \frac {\fD \Nd\Ns}{\fs} \right) & J_0 \! \left(2 \pi \frac {\fD \left( \Nd\Ns + 1 \right)}{\fs} \right) & \!\! \cdots \!\! & J_0 \! \left(2 \pi \frac {\fD \left( \Nd\Ns + \Ns - 1 \right)}{\fs} \right) \\
\!\! J_0 \! \left(2 \pi \frac {\fD \left(\Nd\Ns - 1 \right)}{\fs} \right) & J_0 \! \left(2 \pi \frac {\fD \Nd\Ns}{\fs} \right) & \!\! \cdots \!\! & J_0 \! \left(2 \pi \frac {\fD \left( \Nd\Ns + \Ns - 2 \right)}{\fs} \right) \\
\!\! \vdots & \vdots & \!\! \ddots \!\! & \vdots \\
\!\! J_0 \! \left(2 \pi \frac {\fD \left( \Nd \Ns - \Ns + 1 \right)}{\fs} \right) & J_0 \! \left(2 \pi \frac {\fD \left( \Nd \Ns - \Ns + 2 \right)}{\fs} \right) & \!\! \cdots \!\! & J_0 \! \left(2 \pi \frac {\fD \Nd\Ns}{\fs} \right) \end{array} \right]
\end{equation}
\begin{equation}
\label{e:PI_F}
\mathbf{\Pi}_{\textnormal{F}} = \left[ \begin{array}{cccc} 1 & \textnormal{si} \left(2 \pi \tau_{\textnormal{max}} \Delta F \right) & \cdots & \textnormal{si} \left(2 \pi \tau_{\textnormal{max}} \Delta F \left( \Ns - 1 \right) \right) \\
\textnormal{si} \left(2 \pi \tau_{\textnormal{max}} \Delta F \right) & 1 & \cdots & \textnormal{si} \left(2 \pi \tau_{\textnormal{max}} \Delta F \left( \Ns - 2 \right) \right) \\
\vdots & \vdots & \ddots & \vdots \\
\textnormal{si} \left(2 \pi \tau_{\textnormal{max}} \Delta F \left( \Ns - 1 \right) \right) & \textnormal{si} \left(2 \pi \tau_{\textnormal{max}} \Delta F \left( \Ns - 2 \right) \right) & \cdots & 1 \end{array} \right]
\end{equation}
\end{figure*}
\noindent where $E\{|h|^2\}$ is the variance of the channel coefficients, $\otimes$ denotes the Kronecker product, and $\mathbf{\Pi}_{\textnormal{T}} \left( \Nd \right)$ and $\mathbf{\Pi}_{\textnormal{F}}$ are given in~\eqref{e:PI_T} and~\eqref{e:PI_F} on the next page. Here, $J_0(\cdot)$ is the zero-th order Bessel function of the first kind, $\textnormal{si}(x)$ denotes $\textnormal{sin}(x)/x$, and the maximum Doppler frequency is given as $\fD = \fc \cdot v / c$~\cite{DentCroft_LETTERS93}, where $\fc$ is the carrier frequency, $v$ the UT speed, and $c$ the speed of light. 
$\Nd$ denotes the delay (in TTIs) between the PRB, which is used to estimate the channel, and the PRB to which the estimate is applied at the transmitter. This delay will be of particular interest in the context of CSI feedback later, but is set to zero for the moment. The OFDM-symbol-wise mean-square error (MSE) of the obtained channel estimates can now be stated as~\cite{Hoeher_BOOK97,Hoeher_ICASSP97,CastroJohamUtschick_ASILOMAR07,JohamUtschick_ICASSP08}
\begin{multline}
\label{e:MSE_CSIR}
\MSE^{\textnormal{CSIR}} = \DIAG \left( E\left\{ \left( \tilde{\Bh}^{[t]} - \Bh^{[t]} \right) \left(\tilde{\Bh}^{[t]} - \Bh^{[t]} \right)^H \right\} \right) \\
= E\left\{ \left| h \right|^2 \right\} \Bone \SPA - \\
\DIAG \left( \mathbf{\Phi}_{\textnormal{hh}} \left( 0 \right) \BS^H \inv{ \BS \mathbf{\Phi}_{\textnormal{hh}} \left( 0 \right) \BS^H\!\!+\!\! \sigma_{\textnormal{p}}^2 \BI} \BS \mathbf{\Phi}_{\textnormal{hh}} \left( 0 \right) \right).
\end{multline}

Typically, the MSE of the outer OFDM symbols is worse than that between pilot positions. Since the badly estimated OFDM symbols have a dominant effect on the overall transmission, we calculate a representative value for $\sigma_{\textnormal{DL,UT}}^2$ as
\begin{equation}
\label{e:SIGMA_DL_UE}
\sigma_{\textnormal{DL,UT}}^2 = \frac {\MAX \left( \MSE^{\textnormal{CSIR}} \right)} {E\left\{ \left| h \right|^2 \right\} - \MAX \left( \MSE^{\textnormal{CSIR}} \right)}.
\end{equation}

\subsection{Channel Prediction and CSI Feedback}
As depicted in Fig.~\ref{fig:csifeedback}, the pilots received at the UT side are also used to obtain a channel estimate $\bar{\Bh}^{[t+\Nd]}$ that predicts the channel $\Nd$ TTIs into the future, assuming the feedback process itself consumes $\Nd$ TTIs.
The channel prediction is achieved by the modified MMSE filter
\begin{equation}
\label{e:CSIR_MMSEFILTER_PREDICTION}
\BG_{\textnormal{P}} = \mathbf{\Phi}_{\textnormal{hh}} \left( \Nd \right) \BS^H \inv{ \BS \mathbf{\Phi}_{\textnormal{hh}} \left( 0 \right) \BS^H + \sigma_{\textnormal{p}}^2 \BI}.
\end{equation}

The (again OFDM-symbol-wise) MSE between the predicted channel estimate and the actual channel in the corresponding \ac{TTI} of interest is given as
\begin{multline}
\label{e:MSE_CSIR_PREDICTED}
\MSE^{\textnormal{CSIR,P}} = \DIAG \left( E\left\{ \left( \tilde{\Bh}^{[t]} - \Bh^{[t]} \right) \left(\tilde{\Bh}^{[t]} - \Bh^{[t]} \right)^H \right\} \right) \\
= E\left\{ \left| h \right|^2 \right\} \Bone \SPA - \\
\DIAG \left( \mathbf{\Phi}_{\textnormal{hh}} \! \left( \Nd \right) \BS^H \inv{ \BS \mathbf{\Phi}_{\textnormal{hh}} \! \left( 0 \right) \BS^H \!\! + \!\! \sigma_{\textnormal{p}}^2 \BI} \BS \mathbf{\Phi}_{\textnormal{hh}} \! \left( \Nd \right) \right),
\end{multline}

\noindent which is equivalent to~\eqref{e:MSE_CSIR}, except that CSI feedback delay $\Nd$ is now taken into account. 

\subsection{Redundant CSI Quantization in each TTI}
Our work considers two CSI quantization approaches, which give a lower and upper bound for the performance of a practical system. In the first case, we assume that in each \ac{TTI} $t$, the channel prediction $\bar{\Bh}^{[t+\Nd]}$ is quantized independently of previous estimates and fed back to the BS side. As in~\cite{CastroJohamUtschick_ASILOMAR07, JohamUtschick_ICASSP08}, we assume that a decorrelation filter $\BV^H$ is applied to $\bar{\Bh}^{[t+\Nd]}$, such that we obtain a vector of $\Nrank$ uncorrelated Gaussian quantities. Filter matrix $\BV \in \mathbb{C}^{[\Ns\Nc \times \Nrank]}$ is obtained through an Eigenvalue decomposition of the signal covariance at the output of the MMSE predictor, i.e.
\begin{eqnarray}
\mathbf{\Phi}_{\bar{\textnormal{h}}\bar{\textnormal{h}}} &\!\!\!\!=\!\!\!\!& E \left\{\bar{\Bh}^{[t]}\left(\bar{\Bh}^{[t]}\right)^H \right\} \\ &\!\!\!\!=\!\!\!\!& \mathbf{\Phi}_{\textnormal{hh}} \! \left( \Nd \right) \BS^H \! \inv{ \BS \mathbf{\Phi}_{\textnormal{hh}} \! \left( 0 \right) \BS^H \!\! + \!\! \sigma_{\textnormal{p}}^2 \BI} \BS \mathbf{\Phi}_{\textnormal{hh}} \! \left( \Nd \right) \\
&\!\!\!\!=\!\!\!\!& \BU \mathbf{\Sigma} \BU^H,
\end{eqnarray}

\noindent after which $\BV$ is chosen such that it contains the $\Nrank$ column vectors from $\BU$ that correspond to the strongest Eigenvalues on the diagonal of $\mathbf{\Sigma}$. The rank-reduced channel estimates are quantized, leading to an introduction of quantization noise $\Bn_{\textnormal{q}} \sim \mathcal{N}_{\mathbb{C}}(\Bnull, \Pqq)$. Then, they are fed back to the transmitter through an error-free link, where a multiplication with $\BV$ yields the vector of channel estimates $\hat{\Bh}$. We consider a practical quantization approach~\cite{LindeGray_IEEETRANS80}, where an overall number of $\Nb$ bits is equally invested into each of the $\Nrank$ decorrelated channel estimates, and one bit per real dimension is lost w.r.t. the rate-distortion bound~\cite{Cover.Thomas.1991}. The quantization noise inherent in the feedback can now be stated as
\begin{equation}
\label{e:QUANT_NOISE}
\Pqq = 2^{-\MAX\left( 0, \frac {\Nb}{\Nrank} - 2 \right)} \BV \BV^H \mathbf{\Phi}_{\bar{\textnormal{h}}\bar{\textnormal{h}}} \BV \BV^H.
\end{equation}

Finally, the MSE of the predicted channel at the transmitter side is given as~\cite{CastroJohamUtschick_ASILOMAR07}
\begin{multline}
\label{e:MSE_CSIT}
\MSE^{\textnormal{CSIT}} = \DIAG \left( E\left\{ \left( \hat{\Bh}^{[t]} - \Bh^{[t]} \right) \left(\hat{\Bh}^{[t]} - \Bh^{[t]} \right)^H \right\} \right) \\
= E\left\{ \left| h \right|^2 \right\} \Bone + \DIAG \left( \Pqq - \BV\BV^H \mathbf{\Phi}_{\bar{\textnormal{h}}\bar{\textnormal{h}}} \BV \BV^H \right).
\end{multline}

\subsection{Successive CSI Feedback}
The amount of CSI feedback can be significantly reduced if its correlation in time is exploited. This can be modeled by letting the UTs quantize the channel estimate $\bar{\Bh}^{[t+\Nd]}$ {\em conditioned} on the previous channel estimate $\hat{\Bh}^{[t+\Nd-1]}$ sent to the BS. Hence, we are interested in the {\em conditional covariance}
\begin{multline}
\label{e:CONDITIONAL_COVARIANCE}
\mathbf{\Phi}_{\bar{\Bh}^{[t]}\left(\bar{\Bh}^{[t]}\right)^H \left| \hat{\Bh}^{[t-1]}\right.} = \mathbf{\Phi}_{\bar{\textnormal{h}}\bar{\textnormal{h}}} \\
- E\left\{ \bar{\Bh}^{[t]} \left(\hat{\Bh}^{[t-1]}\right)^H \right\} \inv{E\left\{ \hat{\Bh}\hat{\Bh}^H \right\}} E\left\{ \hat{\Bh}^{[t-1]} \left(\bar{\Bh}^{[t]}\right)^H \right\} \\ = \BU \mathbf{\Sigma} \BU^H,
\end{multline}
\noindent with
\begin{eqnarray}
E\left\{ \bar{\Bh}^{[t]} \left(\hat{\Bh}^{[t-1]}\right)^H \right\} &\!\!\!\!=\!\!\!\!& \beta \BG_{\textnormal{P}} \left(
\BS \mathbf{\Phi}_{\textnormal{hh}} \! \left( -1 \right) \BS^H \!\! + \!\! \sigma_{\textnormal{p}}^2 \BI \right) \BG_{\textnormal{P}}^H \BV \BV^H \nonumber \\
\textnormal{with} \SPA \beta &\!\!\!\!=\!\!\!\!& \sqrt{1-2^{-\MAX \left(\frac{\Nb'}{\Nrank}-2,0\right)}},
\end{eqnarray}

\noindent where $\Nb'$ is the number of CSI feedback bits used in the previous \ac{TTI} (assuming that {\em unconditioned} CSI was fed back at that time). We perform a rank reduction of the conditional covariance given in~\eqref{e:CONDITIONAL_COVARIANCE} as before and calculate the quantization noise under the assumption of the same quantizer as in~\eqref{e:QUANT_NOISE}. From~\eqref{e:MSE_CSIT} we can calculate $\sigma_{\textnormal{DL,BS}}^2$, again based on the assumption that the overall performance is dominated by the OFDM symbols for which channel estimation is worst:
\begin{equation}
\label{e:SIGMA_DL_BS}
\sigma_{\textnormal{DL,BS}}^2 = \frac {\MAX \left( \MSE^{\textnormal{CSIT}} \right) - \MAX \left( \MSE^{\textnormal{CSIR}} \right)} {E\left\{ \left| h \right|^2 \right\} - \MAX \left( \MSE^{\textnormal{CSIR}} \right)}.
\end{equation}

For successive CSI feedback, we can now consider one TTI of unconditioned CSI feedback using $\Nb'$ bits, followed by one TTI of successive CSI feedback with $\Nb$ bits. If we now adjust $\Nb'$ such that the resulting MSE is the same, we obtain the {\em steady-state performance} of a scheme continuously using successive feedback with $\Nb$ bits after an initialization phase.

\medskip

The noise term $\sigma_{\textnormal{UL,UT}}^2$, which reflects the impact of channel estimation error at the receiver in the \ac{UL}, can be calculated using the same methodology presented in this section, but considering only the channel estimation part. 

\section{Joint UL/DL Optimization}
\label{sec:joint.rate.region}

Currently, the asymmetric operation of mobile communication systems is only considered by the radio link control, e.\,g., \ac{RRM} and quality of service. Nonetheless, the joint view of \ac{UL} and \ac{DL} can also be applied to the physical layer and reflects the increasing demand for cross-layer optimization. An example is the adjustment of physical layer parameters in order to adapt to the demands implied by higher layers.

\medskip

To evaluate the trade-off between \ac{UL} and \ac{DL}, we need to take into account the overhead connected to pilots and CSI feedback, and hence find expressions for the {\em net} sum rate in \ac{UL} and \ac{DL}. For the \ac{UL}, this is given as
\begin{equation}
\label{e:NET_RATE_UL}
R_{\textnormal{UL}}^{\textnormal{net}} = \frac{ R_{\textnormal{UL}} \cdot \overbrace{\Ns\Nc \left(1-K \cdot \rho_{\textnormal{UL}} \right)}^{\textnormal{Sum rate without pilot overhead}} \!\!\! - \!\! \overbrace{\Nb \cdot \NBS \cdot K}^{\textnormal{CSI feedback overhead}}}{\Ns\Nc}.
\end{equation}
Equation~\eqref{e:NET_RATE_UL} considers the overall rate in a PRB (without pilot symbols), and subtracts the rate required for CSI feedback. The overall pilot effort is the product of pilot density $p_{\textnormal{UL}}$ and number of UTs $K$, since the BS has to be able to distinguish all terminals based on orthogonal pilot sequences. For the CSI feedback effort, on the other hand, we have to consider that $\Nb$ bits are required for all $\NBS \cdot K$ spatial coefficients of the MIMO channel.

\medskip

The net sum rate in the DL depends on whether spatial multiplexing is performed or TDM with random precoding vectors. In the former case, we can state
\begin{equation}
\label{e:NET_RATE_DL}
R_{\textnormal{DL}}^{\textnormal{net}} = R_{\textnormal{DL}} \left(1 - \left( \NBS + K \right) \cdot \rho_{\textnormal{DL}} \right),
\end{equation}

\noindent as we need one pilot for each of the $\NBS$ BS antennas (required for channel estimation connected to {\em CSI feedback} at the UT side), as well as one pilot for each UT-specific stream (required for channel estimation connected to {\em data decoding} at the UT side)~\cite{Fettweis.etal.ICASSP.2010}. In the case of TDM with random precoding, this increases to
\begin{equation}
\label{e:NET_RATE_DL_TDM}
R_{\textnormal{DL}}^{\textnormal{net}} = R_{\textnormal{DL}} \left(1 - \rho_{\textnormal{DL}} \right),
\end{equation}

\noindent as the transmission is only performed to one UT at a time, and this UT only needs to estimate the effective channel as a result of random precoding.

\medskip

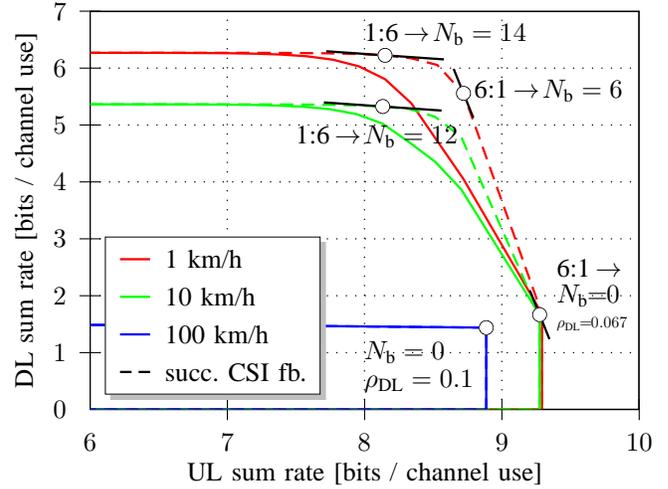
\begin{figure}
      \begingroup
\unitlength=1mm
\psset{xunit=18.25000mm, yunit=7.57143mm, linewidth=0.1mm}
\psset{arrowsize=2pt 3, arrowlength=1.4, arrowinset=.4}\psset{axesstyle=frame}
\begin{pspicture}(5.45205, -1.32075)(10.00000, 7.00000)
\psaxes[subticks=0, labels=all, xsubticks=1, ysubticks=1, Ox=6, Oy=0, Dx=1, Dy=1]{-}(6.00000, 0.00000)(6.00000, 0.00000)(10.00020, 7.00020)%
\multips(7.00000, 0.00000)(1.00000, 0.0){3}{\psline[linecolor=black, linestyle=dotted, linewidth=0.2mm](0, 0)(0, 7.00000)}
\multips(6.00000, 1.00000)(0, 1.00000){6}{\psline[linecolor=black, linestyle=dotted, linewidth=0.2mm](0, 0)(4.00000, 0)}
\rput[b](8.00000, -1.32075){UL sum rate [bits / channel use]}
\psclip{\psframe(6.00000, 0.00000)(10.00000, 7.00000)}
\psline[linecolor=red, plotstyle=curve, linestyle=solid, dotscale=1.5 1.5, showpoints=false, linewidth=0.3mm](0.00000, 1.41047)(0.00000, 1.37577)(0.00000, 1.35890)(0.00000, 1.33669)(0.00000, 1.27300)(0.00000, 0.00000)(9.29261, 0.00000)(9.29261, 1.27300)(9.29261, 1.33669)(9.29261, 1.35890)(9.29261, 1.37577)(9.29261, 1.41047)(9.29261, 1.55251)(9.29261, 1.55629)(9.29261, 1.63581)(9.29261, 1.67142)(9.29261, 1.67844)(9.29261, 1.68481)(8.72118, 4.04869)(8.34023, 5.38379)(8.14975, 5.80542)(7.95927, 6.03114)(7.76880, 6.14847)(7.57832, 6.20836)(7.38784, 6.23862)(7.19737, 6.25384)(7.00689, 6.26147)(6.91165, 6.26370)(6.62594, 6.26720)(6.43546, 6.26815)(6.19665, 6.26815)(6.16966, 6.26815)(6.14453, 6.26815)(5.88557, 6.26815)(5.79892, 6.26815)(5.27484, 6.26815)(4.63488, 6.26815)(3.31223, 6.26815)(0.56959, 6.26815)(0.00000, 6.26815)(0.00000, 1.68481)(0.00000, 1.67844)(0.00000, 1.67142)(0.00000, 1.63581)(0.00000, 1.55629)(0.00000, 1.55251)(0.00000, 1.41047)
\psline[linecolor=red, plotstyle=curve, linestyle=dashed, dotscale=1.5 1.5, showpoints=false, linewidth=0.3mm](0.00000, 1.41047)(0.00000, 1.37577)(0.00000, 1.35890)(0.00000, 1.33669)(0.00000, 1.27300)(0.00000, 0.00000)(9.29261, 0.00000)(9.29261, 1.27300)(9.29261, 1.33669)(9.29261, 1.35890)(9.29261, 1.37577)(9.29261, 1.41047)(9.29261, 1.55251)(9.29261, 1.55629)(9.29261, 1.63581)(9.29261, 1.67142)(9.29261, 1.67844)(9.29261, 1.68481)(8.72118, 5.55676)(8.53070, 6.05427)(8.34023, 6.17363)(8.14975, 6.22444)(7.95927, 6.24755)(7.76880, 6.25852)(7.57832, 6.26387)(7.38784, 6.26650)(7.19737, 6.26781)(7.00689, 6.26846)(6.91165, 6.26865)(6.62594, 6.26895)(6.43546, 6.26903)(6.19665, 6.26903)(6.16966, 6.26903)(6.14453, 6.26903)(5.88557, 6.26903)(5.79892, 6.26903)(5.27484, 6.26903)(4.63488, 6.26903)(3.31223, 6.26903)(0.56959, 6.26903)(0.00000, 6.26903)(0.00000, 1.68481)(0.00000, 1.67844)(0.00000, 1.67142)(0.00000, 1.63581)(0.00000, 1.55629)(0.00000, 1.55251)(0.00000, 1.41047)
\psline[linecolor=black, plotstyle=curve, linestyle=solid, dotscale=1.5 1.5, showpoints=false, linewidth=0.3mm](7.72118, 6.29587)(8.57832, 6.15301)
\psline[linecolor=black, plotstyle=curve, linestyle=solid, dotscale=1.5 1.5, showpoints=false, linewidth=0.3mm](8.14975, 6.22444)(8.14975, 6.22444)
\psline[linecolor=black, plotstyle=curve, dotstyle=o, linestyle=none, dotscale=1.5 1.5, showpoints=true, linewidth=0.3mm](8.14975, 6.22444)(8.14975, 6.22444)
\psline[linecolor=black, plotstyle=curve, linestyle=solid, dotscale=1.5 1.5, showpoints=false, linewidth=0.3mm](8.64975, 5.98533)(8.79261, 5.12819)
\psline[linecolor=black, plotstyle=curve, linestyle=solid, dotscale=1.5 1.5, showpoints=false, linewidth=0.3mm](8.72118, 5.55676)(8.72118, 5.55676)
\psline[linecolor=black, plotstyle=curve, dotstyle=o, linestyle=none, dotscale=1.5 1.5, showpoints=true, linewidth=0.3mm](8.72118, 5.55676)(8.72118, 5.55676)
\psline[linecolor=green, plotstyle=curve, linestyle=solid, dotscale=1.5 1.5, showpoints=false, linewidth=0.3mm](0.00000, 1.39641)(0.00000, 1.36478)(0.00000, 1.34033)(0.00000, 1.32281)(0.00000, 1.26222)(0.00000, 0.00000)(9.27573, 0.00000)(9.27573, 1.26222)(9.27573, 1.32281)(9.27573, 1.34033)(9.27573, 1.36478)(9.27573, 1.39641)(9.27573, 1.53617)(9.27573, 1.54195)(9.27573, 1.61953)(9.27573, 1.65412)(9.27573, 1.65955)(9.27573, 1.66663)(8.70430, 3.86830)(8.51383, 4.35623)(8.32335, 4.70005)(8.13287, 5.02306)(7.94240, 5.19098)(7.75192, 5.27680)(7.56144, 5.32021)(7.37097, 5.34205)(7.18049, 5.35300)(6.99002, 5.35848)(6.89478, 5.36009)(6.60906, 5.36260)(6.41859, 5.36329)(6.17745, 5.36329)(6.15375, 5.36329)(6.11951, 5.36329)(5.86847, 5.36329)(5.78499, 5.36329)(5.25827, 5.36329)(4.61892, 5.36329)(3.29844, 5.36329)(0.56243, 5.36329)(0.00000, 5.36329)(0.00000, 1.66663)(0.00000, 1.65955)(0.00000, 1.65412)(0.00000, 1.61953)(0.00000, 1.54195)(0.00000, 1.53617)(0.00000, 1.39641)
\psline[linecolor=green, plotstyle=curve, linestyle=dashed, dotscale=1.5 1.5, showpoints=false, linewidth=0.3mm](0.00000, 1.39641)(0.00000, 1.36478)(0.00000, 1.34033)(0.00000, 1.32281)(0.00000, 1.26222)(0.00000, 0.00000)(9.27573, 0.00000)(9.27573, 1.26222)(9.27573, 1.32281)(9.27573, 1.34033)(9.27573, 1.36478)(9.27573, 1.39641)(9.27573, 1.53617)(9.27573, 1.54195)(9.27573, 1.61953)(9.27573, 1.65412)(9.27573, 1.65955)(9.27573, 1.66663)(8.70430, 4.79069)(8.51383, 5.14536)(8.32335, 5.28221)(8.13287, 5.32533)(7.94240, 5.34521)(7.75192, 5.35473)(7.56144, 5.35939)(7.37097, 5.36169)(7.18049, 5.36283)(6.99002, 5.36340)(6.89478, 5.36357)(6.60906, 5.36383)(6.41859, 5.36390)(6.17745, 5.36390)(6.15375, 5.36390)(6.11951, 5.36390)(5.86847, 5.36390)(5.78499, 5.36390)(5.25827, 5.36390)(4.61892, 5.36390)(3.29844, 5.36390)(0.56243, 5.36390)(0.00000, 5.36390)(0.00000, 1.66663)(0.00000, 1.65955)(0.00000, 1.65412)(0.00000, 1.61953)(0.00000, 1.54195)(0.00000, 1.53617)(0.00000, 1.39641)
\psline[linecolor=black, plotstyle=curve, linestyle=solid, dotscale=1.5 1.5, showpoints=false, linewidth=0.3mm](7.70430, 5.39675)(8.56144, 5.25390)
\psline[linecolor=black, plotstyle=curve, linestyle=solid, dotscale=1.5 1.5, showpoints=false, linewidth=0.3mm](8.13287, 5.32533)(8.13287, 5.32533)
\psline[linecolor=black, plotstyle=curve, dotstyle=o, linestyle=none, dotscale=1.5 1.5, showpoints=true, linewidth=0.3mm](8.13287, 5.32533)(8.13287, 5.32533)
\psline[linecolor=black, plotstyle=curve, linestyle=solid, dotscale=1.5 1.5, showpoints=false, linewidth=0.3mm](9.20430, 2.09520)(9.34716, 1.23806)
\psline[linecolor=black, plotstyle=curve, linestyle=solid, dotscale=1.5 1.5, showpoints=false, linewidth=0.3mm](9.27573, 1.66663)(9.27573, 1.66663)
\psline[linecolor=black, plotstyle=curve, dotstyle=o, linestyle=none, dotscale=1.5 1.5, showpoints=true, linewidth=0.3mm](9.27573, 1.66663)(9.27573, 1.66663)
\psline[linecolor=blue, plotstyle=curve, linestyle=solid, dotscale=1.5 1.5, showpoints=false, linewidth=0.3mm](0.00000, 1.25060)(0.00000, 1.24652)(0.00000, 1.05282)(0.00000, 1.01270)(0.00000, 0.98873)(0.00000, 0.92586)(0.00000, 0.00000)(8.88696, 0.00000)(8.88696, 0.92586)(8.88696, 0.98873)(8.88696, 1.01270)(8.88696, 1.05282)(8.88696, 1.24652)(8.88696, 1.25060)(8.88696, 1.35345)(8.88696, 1.37249)(8.88696, 1.40594)(8.88696, 1.42528)(8.88696, 1.43920)(7.17267, 1.48009)(6.98219, 1.48322)(6.79172, 1.48478)(6.60124, 1.48556)(6.50600, 1.48579)(6.22029, 1.48615)(6.02981, 1.48624)(5.72971, 1.48624)(5.62254, 1.48624)(5.58818, 1.48624)(5.41525, 1.48624)(5.32312, 1.48624)(5.00324, 1.48624)(4.39303, 1.48624)(3.12880, 1.48624)(0.47852, 1.48624)(0.00000, 1.48624)(0.00000, 1.43920)(0.00000, 1.42528)(0.00000, 1.40594)(0.00000, 1.37249)(0.00000, 1.35345)(0.00000, 1.25060)
\psline[linecolor=blue, plotstyle=curve, linestyle=dashed, dotscale=1.5 1.5, showpoints=false, linewidth=0.3mm](0.00000, 1.25060)(0.00000, 1.24652)(0.00000, 1.05282)(0.00000, 1.01270)(0.00000, 0.98873)(0.00000, 0.92586)(0.00000, 0.00000)(8.88696, 0.00000)(8.88696, 0.92586)(8.88696, 0.98873)(8.88696, 1.01270)(8.88696, 1.05282)(8.88696, 1.24652)(8.88696, 1.25060)(8.88696, 1.35345)(8.88696, 1.37249)(8.88696, 1.40594)(8.88696, 1.42528)(8.88696, 1.43920)(6.98219, 1.48201)(6.79172, 1.48418)(6.60124, 1.48526)(6.50600, 1.48558)(6.22029, 1.48607)(6.02981, 1.48621)(5.72971, 1.48621)(5.62254, 1.48621)(5.58818, 1.48621)(5.41525, 1.48621)(5.32312, 1.48621)(5.00324, 1.48621)(4.39303, 1.48621)(3.12880, 1.48621)(0.47852, 1.48621)(0.00000, 1.48621)(0.00000, 1.43920)(0.00000, 1.42528)(0.00000, 1.40594)(0.00000, 1.37249)(0.00000, 1.35345)(0.00000, 1.25060)
\psline[linecolor=black, plotstyle=curve, dotstyle=o, linestyle=none, dotscale=1.5 1.5, showpoints=true, linewidth=0.3mm](8.88696, 1.43920)(8.88696, 1.43920)
\rput[l](8.00000, 6.60000){1:6 $\!\rightarrow\! \Nb=14$}
\rput[l](8.80000, 5.60000){6:1 $\!\rightarrow\! \Nb=6$}
\rput[l](7.50000, 4.80000){1:6 $\!\rightarrow\! \Nb=12$}
\rput[l](9.40000, 2.50000){6:1 $\!\rightarrow\!$}
\rput[l](9.40000, 2.00000){$\Nb\!\!=\!\!0$}
\rput[l](9.40000, 1.50000){\scalebox{0.6}{$\rho_{\textnormal{DL}}\!\!=\!\!0.067$}}
\rput[l](8.00000, 1.00000){$\Nb=0$}
\rput[l](8.00000, 0.50000){$\rho_{\textnormal{DL}}=0.1$}
\endpsclip
\rput[t]{90}(5.45205, 3.50000){DL sum rate [bits / channel use]}
\psframe[linecolor=black, fillstyle=solid, fillcolor=white, shadowcolor=lightgray, shadowsize=1mm, shadow=true](6.10959, 0.26415)(7.67945, 3.03774)
\rput[l](6.54795, 2.64151){1 km/h}
\psline[linecolor=red, linestyle=solid, linewidth=0.3mm](6.21918, 2.64151)(6.43836, 2.64151)
\rput[l](6.54795, 1.98113){10 km/h}
\psline[linecolor=green, linestyle=solid, linewidth=0.3mm](6.21918, 1.98113)(6.43836, 1.98113)
\rput[l](6.54795, 1.32075){100 km/h}
\psline[linecolor=blue, linestyle=solid, linewidth=0.3mm](6.21918, 1.32075)(6.43836, 1.32075)
\rput[l](6.54795, 0.66038){succ. CSI fb.}
\psline[linecolor=black, linestyle=dashed, linewidth=0.3mm](6.21918, 0.66038)(6.43836, 0.66038)
\end{pspicture}
\endgroup
 
      \caption{Joint UL/DL rate region for an example channel with $\NBS=K=4$ and the corresponding optimal choice of parameters for different target UL/DL rate ratios and terminal velocities.}
      \label{fig:uldl_regions}
    \end{figure}

In our work, we perform a brute-force search over various concrete pilot sequences $\BS$ (yielding different densities $\rho_{\textnormal{UL}}$, $\rho_{\textnormal{DL}}$) and different CSI feedback extents $\Nb$, which allows us to compute convex joint \ac{UL}/\ac{DL} rate regions as given for an example channel realization in Fig.~\ref{fig:uldl_regions}. Each point on the surface of such a rate region is connected to a Pareto-optimal set of parameters $\rho_{\textnormal{UL}}$, $\rho_{\textnormal{DL}}$ and $\Nb$, and constitutes the optimum w.r.t. a certain {\em weighted \ac{UL}/\ac{DL} sum-rate} optimization. In the example, the case of $v=1$ km/h and a strong focus on the \ac{UL} leads to a choice of $\Nb=14$, while $\Nb=6$ is preferable in the case of weighting \ac{UL} and \ac{DL} $1:6$. For $v=100$ km/h, we can see that regardless of \ac{UL}/\ac{DL} weights it is optimal to set $\Nb=0$ and operate the DL in TDM mode with an increased pilot density $\rho_{\textnormal{DL}}=0.1$.

\section{Results}\label{sec:results}
    
    In this section, we present the best choice of pilot densities and CSI feedback quantity as a function of target \ac{UL}/\ac{DL} rate ratio and terminal velocity. In general, we observe a scenario with $\NBS=K=4$ for $f_c=2.6$ GHz, and perform Monte-Carlo simulations with a large number of independent channel realizations in \ac{UL} and \ac{DL}, where all channel coefficients are uncorrelated in space, have zero-mean and are of unit variance $E\{|h|^2\}=1$. All noise terms are set to $\sigma_{\textnormal{UL}}^2\!=\!\sigma_{\textnormal{DL}}^2\!=\!\sigma_{\textnormal{p}}^2\!=\!0.1$, the sum transmit power at both BS and UT side is set to $1$, and the maximum delay spread is $\tau_{\textnormal{max}}=1 \mu$s, which can be seen as a worst-case delay in an urban scenario with rich scattering. For CSI feedback, $\Nrank=2$ is chosen empirically.
    
    \medskip
    \begin{figure*}
\centerline{\subfigure[Optimal number of feedback bits.]{\begingroup
\unitlength=1mm
\psset{xunit=29.20000mm, yunit=2.65000mm, linewidth=0.1mm}
\psset{arrowsize=2pt 3, arrowlength=1.4, arrowinset=.4}\psset{axesstyle=frame}
\begin{pspicture}(-0.34247, -3.77358)(2.50000, 20.00000)
\psaxes[subticks=10, xlogBase=10, logLines=x, labels=all, xsubticks=1, ysubticks=1, Ox=0, Oy=0, Dx=1, Dy=5]{-}(0.00000, 0.00000)(0.00000, 0.00000)(2.50020, 20.00020)%
\multips(0.00000, 5.00000)(0, 5.00000){3}{\psline[linecolor=black, linestyle=dotted, linewidth=0.2mm](0, 0)(2.50000, 0)}
\rput[b](1.25000, -3.77358){UE speed [km/h]}
\psclip{\psframe(0.00000, 0.00000)(2.50000, 20.00000)}
\psline[linecolor=red, plotstyle=curve, linestyle=solid, dotscale=1.5 1.5, showpoints=false, linewidth=0.3mm](0.00000, 18.12500)(0.25001, 18.05000)(0.50000, 18.00000)(0.75000, 18.00000)(1.00000, 17.90000)(1.25000, 15.05000)(1.50000, 11.20000)(1.75000, 3.02500)(2.00000, 0.45000)(2.25000, 0.00000)(2.50000, 0.00000)
\psline[linecolor=red, plotstyle=curve, dotstyle=o, linestyle=none, dotscale=1.5 1.5, showpoints=true, linewidth=0.3mm](0.00000, 18.12500)(1.00000, 17.90000)(2.00000, 0.45000)
\psline[linecolor=red, plotstyle=curve, linestyle=dashed, dotscale=1.5 1.5, showpoints=false, linewidth=0.3mm](0.00000, 15.97500)(0.25001, 16.00000)(0.50000, 15.97500)(0.75000, 15.92500)(1.00000, 15.55000)(1.25000, 13.57500)(1.50000, 10.80000)(1.75000, 3.12500)(2.00000, 0.45000)(2.25000, 0.00000)(2.50000, 0.00000)
\psline[linecolor=red, plotstyle=curve, dotstyle=o, linestyle=none, dotscale=1.5 1.5, showpoints=true, linewidth=0.3mm](0.00000, 15.97500)(1.00000, 15.55000)(2.00000, 0.45000)
\psline[linecolor=green, plotstyle=curve, linestyle=solid, dotscale=1.5 1.5, showpoints=false, linewidth=0.3mm](0.00000, 14.00000)(0.25001, 14.00000)(0.50000, 14.00000)(0.75000, 13.67500)(1.00000, 11.60000)(1.25000, 7.60000)(1.50000, 2.10000)(1.75000, 0.30000)(2.00000, 0.00000)(2.25000, 0.00000)(2.50000, 0.00000)
\psline[linecolor=green, plotstyle=curve, dotstyle=o, linestyle=none, dotscale=1.5 1.5, showpoints=true, linewidth=0.3mm](0.00000, 14.00000)(1.00000, 11.60000)(2.00000, 0.00000)
\psline[linecolor=green, plotstyle=curve, linestyle=dashed, dotscale=1.5 1.5, showpoints=false, linewidth=0.3mm](0.00000, 11.95000)(0.25001, 11.97500)(0.50000, 11.95000)(0.75000, 11.90000)(1.00000, 11.35000)(1.25000, 8.52500)(1.50000, 3.10000)(1.75000, 0.30000)(2.00000, 0.00000)(2.25000, 0.00000)(2.50000, 0.00000)
\psline[linecolor=green, plotstyle=curve, dotstyle=o, linestyle=none, dotscale=1.5 1.5, showpoints=true, linewidth=0.3mm](0.00000, 11.95000)(1.00000, 11.35000)(2.00000, 0.00000)
\psline[linecolor=blue, plotstyle=curve, linestyle=solid, dotscale=1.5 1.5, showpoints=false, linewidth=0.3mm](0.00000, 0.00000)(0.25001, 0.00000)(0.50000, 0.00000)(0.75000, 0.00000)(1.00000, 0.00000)(1.25000, 0.00000)(1.50000, 0.00000)(1.75000, 0.00000)(2.00000, 0.00000)(2.25000, 0.00000)(2.50000, 0.00000)
\psline[linecolor=blue, plotstyle=curve, dotstyle=o, linestyle=none, dotscale=1.5 1.5, showpoints=true, linewidth=0.3mm](0.00000, 0.00000)(1.00000, 0.00000)(2.00000, 0.00000)
\psline[linecolor=blue, plotstyle=curve, linestyle=dashed, dotscale=1.5 1.5, showpoints=false, linewidth=0.3mm](0.00000, 2.45000)(0.25001, 2.32500)(0.50000, 1.75000)(0.75000, 1.25000)(1.00000, 0.45000)(1.25000, 0.00000)(1.50000, 0.00000)(1.75000, 0.00000)(2.00000, 0.00000)(2.25000, 0.00000)(2.50000, 0.00000)
\psline[linecolor=blue, plotstyle=curve, dotstyle=o, linestyle=none, dotscale=1.5 1.5, showpoints=true, linewidth=0.3mm](0.00000, 2.45000)(1.00000, 0.45000)(2.00000, 0.00000)
\rput(0.50000,11.00000){\cnodeput[fillstyle=solid, fillcolor=lightgray](0,0){xx}{\scalebox{0.7}{A}}}
\rput(1.90000,5.00000){\cnodeput[fillstyle=solid, fillcolor=lightgray](0,0){xx}{\scalebox{0.7}{B}}}
\rput(0.20000,2.00000){\cnodeput[fillstyle=solid, fillcolor=lightgray](0,0){xx}{\scalebox{0.7}{C}}}
\endpsclip
\rput[t]{90}(-0.34247, 10.00000){opt. no. CSI fb. bits per coeff.}
\psframe[linecolor=black, fillstyle=solid, fillcolor=white, shadowcolor=lightgray, shadowsize=1mm, shadow=true](1.47260, 11.32075)(2.43151, 19.24528)
\rput[l](1.74658, 18.11321){UL/DL 1:6}
\psline[linecolor=red, linestyle=solid, linewidth=0.3mm](1.54110, 18.11321)(1.67808, 18.11321)
\rput[l](1.74658, 16.22642){UL/DL 1:1}
\psline[linecolor=green, linestyle=solid, linewidth=0.3mm](1.54110, 16.22642)(1.67808, 16.22642)
\rput[l](1.74658, 14.33962){UL/DL 6:1}
\psline[linecolor=blue, linestyle=solid, linewidth=0.3mm](1.54110, 14.33962)(1.67808, 14.33962)
\rput[l](1.74658, 12.45283){succ. CSI fb.}
\psline[linecolor=black, linestyle=dashed, linewidth=0.3mm](1.54110, 12.45283)(1.67808, 12.45283)
\end{pspicture}
\endgroup
 
\label{f:params_vs_speed_nd}} \hfil
\subfigure[Optimal pilot density.]{\begingroup
\unitlength=1mm
\psset{xunit=29.20000mm, yunit=353.33333mm, linewidth=0.1mm}
\psset{arrowsize=2pt 3, arrowlength=1.4, arrowinset=.4}\psset{axesstyle=frame}
\begin{pspicture}(-0.34247, -0.02830)(2.50000, 0.15000)
\psaxes[subticks=10, xlogBase=10, logLines=x, labels=all, xsubticks=1, ysubticks=1, Ox=0, Oy=0, Dx=1, Dy=0.05]{-}(0.00000, 0.00000)(0.00000, 0.00000)(2.50020, 0.15020)%
\multips(0.00000, 0.05000)(0, 0.05000){2}{\psline[linecolor=black, linestyle=dotted, linewidth=0.2mm](0, 0)(2.50000, 0)}
\rput[b](1.25000, -0.02830){UE speed [km/h]}
\psclip{\psframe(0.00000, 0.00000)(2.50000, 0.15000)}
\psline[linecolor=red, plotstyle=curve, linestyle=solid, dotscale=1.5 1.5, showpoints=false, linewidth=0.3mm](0.00000, 0.01667)(0.25001, 0.01667)(0.50000, 0.01667)(0.75000, 0.01667)(1.00000, 0.01667)(1.25000, 0.01667)(1.50000, 0.01667)(1.75000, 0.01667)(2.00000, 0.01667)(2.25000, 0.01688)(2.50000, 0.03333)
\psline[linecolor=red, plotstyle=curve, dotstyle=triangle, dotangle=0, linestyle=none, dotscale=1.5 1.5, showpoints=true, linewidth=0.3mm](0.00000, 0.01667)(1.00000, 0.01667)(2.00000, 0.01667)
\psline[linecolor=red, plotstyle=curve, linestyle=dashed, dotscale=1.5 1.5, showpoints=false, linewidth=0.3mm](0.00000, 0.01667)(0.25001, 0.01667)(0.50000, 0.01667)(0.75000, 0.01667)(1.00000, 0.01667)(1.25000, 0.01667)(1.50000, 0.01667)(1.75000, 0.01667)(2.00000, 0.01667)(2.25000, 0.01688)(2.50000, 0.03333)
\psline[linecolor=red, plotstyle=curve, dotstyle=triangle, dotangle=0, linestyle=none, dotscale=1.5 1.5, showpoints=true, linewidth=0.3mm](0.00000, 0.01667)(1.00000, 0.01667)(2.00000, 0.01667)
\psline[linecolor=red, plotstyle=curve, linestyle=solid, dotscale=1.5 1.5, showpoints=false, linewidth=0.3mm](0.00000, 0.01667)(0.25001, 0.01667)(0.50000, 0.01667)(0.75000, 0.01667)(1.00000, 0.01667)(1.25000, 0.01729)(1.50000, 0.02625)(1.75000, 0.07208)(2.00000, 0.09292)(2.25000, 0.09042)(2.50000, 0.08208)
\psline[linecolor=red, plotstyle=curve, dotstyle=triangle, dotangle=180, linestyle=none, dotscale=1.5 1.5, showpoints=true, linewidth=0.3mm](0.00000, 0.01667)(1.00000, 0.01667)(2.00000, 0.09292)
\psline[linecolor=red, plotstyle=curve, linestyle=dashed, dotscale=1.5 1.5, showpoints=false, linewidth=0.3mm](0.00000, 0.01667)(0.25001, 0.01667)(0.50000, 0.01667)(0.75000, 0.01667)(1.00000, 0.01667)(1.25000, 0.01729)(1.50000, 0.02562)(1.75000, 0.07104)(2.00000, 0.09292)(2.25000, 0.09042)(2.50000, 0.08208)
\psline[linecolor=red, plotstyle=curve, dotstyle=triangle, dotangle=180, linestyle=none, dotscale=1.5 1.5, showpoints=true, linewidth=0.3mm](0.00000, 0.01667)(1.00000, 0.01667)(2.00000, 0.09292)
\psline[linecolor=green, plotstyle=curve, linestyle=solid, dotscale=1.5 1.5, showpoints=false, linewidth=0.3mm](0.00000, 0.01667)(0.25001, 0.01667)(0.50000, 0.01667)(0.75000, 0.01667)(1.00000, 0.01667)(1.25000, 0.01667)(1.50000, 0.01667)(1.75000, 0.01667)(2.00000, 0.01667)(2.25000, 0.01688)(2.50000, 0.03333)
\psline[linecolor=green, plotstyle=curve, dotstyle=triangle, dotangle=0, linestyle=none, dotscale=1.5 1.5, showpoints=true, linewidth=0.3mm](0.00000, 0.01667)(1.00000, 0.01667)(2.00000, 0.01667)
\psline[linecolor=green, plotstyle=curve, linestyle=dashed, dotscale=1.5 1.5, showpoints=false, linewidth=0.3mm](0.00000, 0.01667)(0.25001, 0.01667)(0.50000, 0.01667)(0.75000, 0.01667)(1.00000, 0.01667)(1.25000, 0.01667)(1.50000, 0.01667)(1.75000, 0.01667)(2.00000, 0.01667)(2.25000, 0.01688)(2.50000, 0.03333)
\psline[linecolor=green, plotstyle=curve, dotstyle=triangle, dotangle=0, linestyle=none, dotscale=1.5 1.5, showpoints=true, linewidth=0.3mm](0.00000, 0.01667)(1.00000, 0.01667)(2.00000, 0.01667)
\psline[linecolor=green, plotstyle=curve, linestyle=solid, dotscale=1.5 1.5, showpoints=false, linewidth=0.3mm](0.00000, 0.01667)(0.25001, 0.01667)(0.50000, 0.01667)(0.75000, 0.01729)(1.00000, 0.01792)(1.25000, 0.02667)(1.50000, 0.05667)(1.75000, 0.08771)(2.00000, 0.09542)(2.25000, 0.09042)(2.50000, 0.08208)
\psline[linecolor=green, plotstyle=curve, dotstyle=triangle, dotangle=180, linestyle=none, dotscale=1.5 1.5, showpoints=true, linewidth=0.3mm](0.00000, 0.01667)(1.00000, 0.01792)(2.00000, 0.09542)
\psline[linecolor=green, plotstyle=curve, linestyle=dashed, dotscale=1.5 1.5, showpoints=false, linewidth=0.3mm](0.00000, 0.01667)(0.25001, 0.01667)(0.50000, 0.01667)(0.75000, 0.01667)(1.00000, 0.01729)(1.25000, 0.02062)(1.50000, 0.05167)(1.75000, 0.08771)(2.00000, 0.09542)(2.25000, 0.09042)(2.50000, 0.08208)
\psline[linecolor=green, plotstyle=curve, dotstyle=triangle, dotangle=180, linestyle=none, dotscale=1.5 1.5, showpoints=true, linewidth=0.3mm](0.00000, 0.01667)(1.00000, 0.01729)(2.00000, 0.09542)
\psline[linecolor=blue, plotstyle=curve, linestyle=solid, dotscale=1.5 1.5, showpoints=false, linewidth=0.3mm](0.00000, 0.01667)(0.25001, 0.01667)(0.50000, 0.01667)(0.75000, 0.01667)(1.00000, 0.01667)(1.25000, 0.01667)(1.50000, 0.01667)(1.75000, 0.01667)(2.00000, 0.01667)(2.25000, 0.01688)(2.50000, 0.03333)
\psline[linecolor=blue, plotstyle=curve, dotstyle=triangle, dotangle=0, linestyle=none, dotscale=1.5 1.5, showpoints=true, linewidth=0.3mm](0.00000, 0.01667)(1.00000, 0.01667)(2.00000, 0.01667)
\psline[linecolor=blue, plotstyle=curve, linestyle=dashed, dotscale=1.5 1.5, showpoints=false, linewidth=0.3mm](0.00000, 0.01667)(0.25001, 0.01667)(0.50000, 0.01667)(0.75000, 0.01667)(1.00000, 0.01667)(1.25000, 0.01667)(1.50000, 0.01667)(1.75000, 0.01667)(2.00000, 0.01667)(2.25000, 0.01688)(2.50000, 0.03333)
\psline[linecolor=blue, plotstyle=curve, dotstyle=triangle, dotangle=0, linestyle=none, dotscale=1.5 1.5, showpoints=true, linewidth=0.3mm](0.00000, 0.01667)(1.00000, 0.01667)(2.00000, 0.01667)
\psline[linecolor=blue, plotstyle=curve, linestyle=solid, dotscale=1.5 1.5, showpoints=false, linewidth=0.3mm](0.00000, 0.06937)(0.25001, 0.06937)(0.50000, 0.06937)(0.75000, 0.06917)(1.00000, 0.06917)(1.25000, 0.06917)(1.50000, 0.07646)(1.75000, 0.09083)(2.00000, 0.09542)(2.25000, 0.09042)(2.50000, 0.08208)
\psline[linecolor=blue, plotstyle=curve, dotstyle=triangle, dotangle=180, linestyle=none, dotscale=1.5 1.5, showpoints=true, linewidth=0.3mm](0.00000, 0.06937)(1.00000, 0.06917)(2.00000, 0.09542)
\psline[linecolor=blue, plotstyle=curve, linestyle=dashed, dotscale=1.5 1.5, showpoints=false, linewidth=0.3mm](0.00000, 0.05313)(0.25001, 0.05375)(0.50000, 0.05729)(0.75000, 0.05958)(1.00000, 0.06521)(1.25000, 0.06917)(1.50000, 0.07646)(1.75000, 0.09083)(2.00000, 0.09542)(2.25000, 0.09042)(2.50000, 0.08208)
\psline[linecolor=blue, plotstyle=curve, dotstyle=triangle, dotangle=180, linestyle=none, dotscale=1.5 1.5, showpoints=true, linewidth=0.3mm](0.00000, 0.05313)(1.00000, 0.06521)(2.00000, 0.09542)
\rput(2.30000,0.03000){\cnodeput[fillstyle=solid, fillcolor=lightgray](0,0){xx}{\scalebox{0.7}{D}}}
\rput(1.20000,0.05000){\cnodeput[fillstyle=solid, fillcolor=lightgray](0,0){xx}{\scalebox{0.7}{E}}}
\rput(2.30000,0.07000){\cnodeput[fillstyle=solid, fillcolor=lightgray](0,0){xx}{\scalebox{0.7}{F}}}
\endpsclip
\rput[t]{90}(-0.34247, 0.07500){opt. pilot density}
\psframe[linecolor=black, fillstyle=solid, fillcolor=white, shadowcolor=lightgray, shadowsize=1mm, shadow=true](0.06849, 0.07075)(0.92466, 0.14434)
\rput[l](0.34247, 0.13585){UL/DL 1:6}
\psline[linecolor=red, linestyle=solid, linewidth=0.3mm](0.13699, 0.13585)(0.27397, 0.13585)
\rput[l](0.34247, 0.12170){UL/DL 1:1}
\psline[linecolor=green, linestyle=solid, linewidth=0.3mm](0.13699, 0.12170)(0.27397, 0.12170)
\rput[l](0.34247, 0.10755){UL/DL 6:1}
\psline[linecolor=blue, linestyle=solid, linewidth=0.3mm](0.13699, 0.10755)(0.27397, 0.10755)
\rput[l](0.34247, 0.09340){uplink}
\psline[linecolor=black, linestyle=solid, linewidth=0.3mm](0.13699, 0.09340)(0.27397, 0.09340)
\psline[linecolor=black, linestyle=solid, linewidth=0.3mm](0.13699, 0.09340)(0.27397, 0.09340)
\psdots[linecolor=black, linestyle=solid, linewidth=0.3mm, dotstyle=triangle, dotangle=0, dotscale=1.5 1.5, linecolor=black](0.20548, 0.09340)
\rput[l](0.34247, 0.07925){downlink}
\psline[linecolor=black, linestyle=solid, linewidth=0.3mm](0.13699, 0.07925)(0.27397, 0.07925)
\psline[linecolor=black, linestyle=solid, linewidth=0.3mm](0.13699, 0.07925)(0.27397, 0.07925)
\psdots[linecolor=black, linestyle=solid, linewidth=0.3mm, dotstyle=triangle, dotangle=180, dotscale=1.5 1.5, linecolor=black](0.20548, 0.07925)

\end{pspicture}
\endgroup
 
\label{f:params_vs_speed_pd}}}
\caption{Optimal choice of parameters as a function of UT speed.} \label{f:params_vs_speed}
\end{figure*}
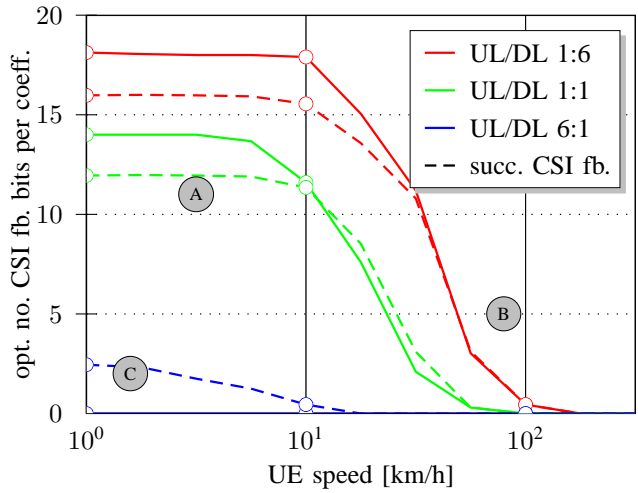
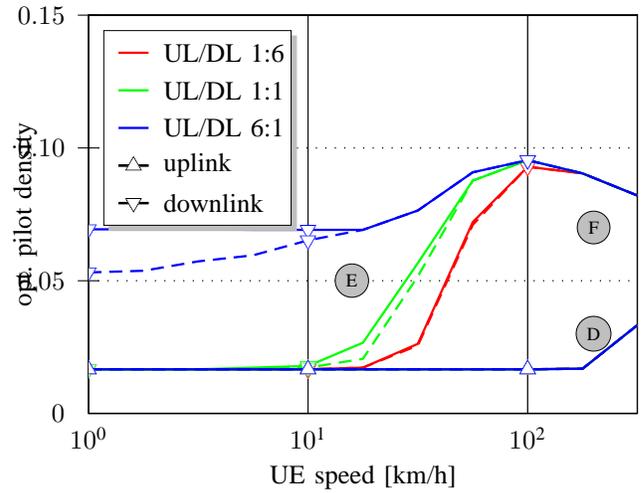

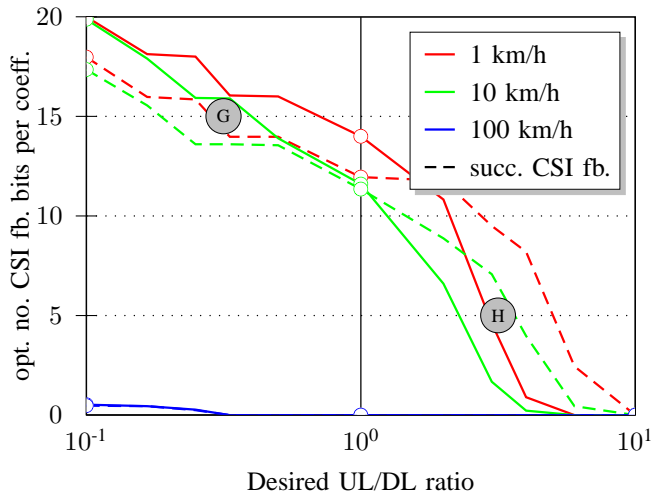
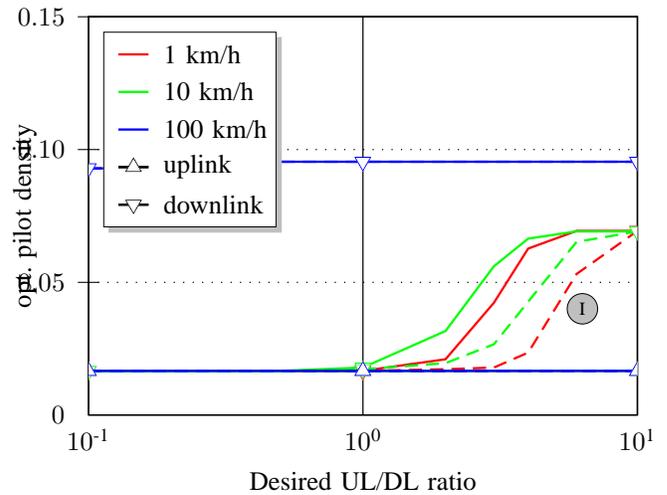
\begin{figure*}
\centerline{\subfigure[Optimal number of feedback bits.]{\begingroup
\unitlength=1mm
\psset{xunit=36.50000mm, yunit=2.65000mm, linewidth=0.1mm}
\psset{arrowsize=2pt 3, arrowlength=1.4, arrowinset=.4}\psset{axesstyle=frame}
\begin{pspicture}(-1.27397, -3.77358)(1.00000, 20.00000)
\psaxes[subticks=10, xlogBase=10, logLines=x, labels=all, xsubticks=1, ysubticks=1, Ox=-1, Oy=0, Dx=1, Dy=5]{-}(-1.00000, 0.00000)(-1.00000, 0.00000)(1.00020, 20.00020)%
\multips(-1.00000, 5.00000)(0, 5.00000){3}{\psline[linecolor=black, linestyle=dotted, linewidth=0.2mm](0, 0)(2.00000, 0)}
\rput[b](0.00000, -3.77358){Desired UL/DL ratio}
\psclip{\psframe(-1.00000, 0.00000)(1.00000, 20.00000)}
\psline[linecolor=red, plotstyle=curve, linestyle=solid, dotscale=1.5 1.5, showpoints=false, linewidth=0.3mm](-1.00000, 20.00000)(-0.77815, 18.12500)(-0.60206, 18.00000)(-0.47712, 16.05000)(-0.30103, 16.00000)(0.00000, 14.00000)(0.30103, 10.82500)(0.47712, 4.60000)(0.60206, 0.90000)(0.77815, 0.00000)(1.00000, 0.00000)
\psline[linecolor=red, plotstyle=curve, dotstyle=o, linestyle=none, dotscale=1.5 1.5, showpoints=true, linewidth=0.3mm](-1.00000, 20.00000)(0.00000, 14.00000)(1.00000, 0.00000)
\psline[linecolor=red, plotstyle=curve, linestyle=dashed, dotscale=1.5 1.5, showpoints=false, linewidth=0.3mm](-1.00000, 17.97500)(-0.77815, 15.97500)(-0.60206, 15.85000)(-0.47712, 13.97500)(-0.30103, 13.97500)(0.00000, 11.95000)(0.30103, 11.77500)(0.47712, 9.50000)(0.60206, 8.20000)(0.77815, 2.45000)(1.00000, 0.00000)
\psline[linecolor=red, plotstyle=curve, dotstyle=o, linestyle=none, dotscale=1.5 1.5, showpoints=true, linewidth=0.3mm](-1.00000, 17.97500)(0.00000, 11.95000)(1.00000, 0.00000)
\psline[linecolor=green, plotstyle=curve, linestyle=solid, dotscale=1.5 1.5, showpoints=false, linewidth=0.3mm](-1.00000, 19.87500)(-0.77815, 17.90000)(-0.60206, 15.92500)(-0.47712, 15.90000)(-0.30103, 13.90000)(0.00000, 11.60000)(0.30103, 6.60000)(0.47712, 1.67500)(0.60206, 0.22500)(0.77815, 0.00000)(1.00000, 0.00000)
\psline[linecolor=green, plotstyle=curve, dotstyle=o, linestyle=none, dotscale=1.5 1.5, showpoints=true, linewidth=0.3mm](-1.00000, 19.87500)(0.00000, 11.60000)(1.00000, 0.00000)
\psline[linecolor=green, plotstyle=curve, linestyle=dashed, dotscale=1.5 1.5, showpoints=false, linewidth=0.3mm](-1.00000, 17.35000)(-0.77815, 15.55000)(-0.60206, 13.60000)(-0.47712, 13.60000)(-0.30103, 13.55000)(0.00000, 11.35000)(0.30103, 8.87500)(0.47712, 7.07500)(0.60206, 3.97500)(0.77815, 0.45000)(1.00000, 0.00000)
\psline[linecolor=green, plotstyle=curve, dotstyle=o, linestyle=none, dotscale=1.5 1.5, showpoints=true, linewidth=0.3mm](-1.00000, 17.35000)(0.00000, 11.35000)(1.00000, 0.00000)
\psline[linecolor=blue, plotstyle=curve, linestyle=solid, dotscale=1.5 1.5, showpoints=false, linewidth=0.3mm](-1.00000, 0.52500)(-0.77815, 0.45000)(-0.60206, 0.27500)(-0.47712, 0.00000)(-0.30103, 0.00000)(0.00000, 0.00000)(0.30103, 0.00000)(0.47712, 0.00000)(0.60206, 0.00000)(0.77815, 0.00000)(1.00000, 0.00000)
\psline[linecolor=blue, plotstyle=curve, dotstyle=o, linestyle=none, dotscale=1.5 1.5, showpoints=true, linewidth=0.3mm](-1.00000, 0.52500)(0.00000, 0.00000)(1.00000, 0.00000)
\psline[linecolor=blue, plotstyle=curve, linestyle=dashed, dotscale=1.5 1.5, showpoints=false, linewidth=0.3mm](-1.00000, 0.47500)(-0.77815, 0.45000)(-0.60206, 0.25000)(-0.47712, 0.00000)(-0.30103, 0.00000)(0.00000, 0.00000)(0.30103, 0.00000)(0.47712, 0.00000)(0.60206, 0.00000)(0.77815, 0.00000)(1.00000, 0.00000)
\psline[linecolor=blue, plotstyle=curve, dotstyle=o, linestyle=none, dotscale=1.5 1.5, showpoints=true, linewidth=0.3mm](-1.00000, 0.47500)(0.00000, 0.00000)(1.00000, 0.00000)
\rput(-0.50000,15.00000){\cnodeput[fillstyle=solid, fillcolor=lightgray](0,0){xx}{\scalebox{0.7}{G}}}
\rput(0.50000,5.00000){\cnodeput[fillstyle=solid, fillcolor=lightgray](0,0){xx}{\scalebox{0.7}{H}}}
\endpsclip
\rput[t]{90}(-1.27397, 10.00000){opt. no. CSI fb. bits per coeff.}
\psframe[linecolor=black, fillstyle=solid, fillcolor=white, shadowcolor=lightgray, shadowsize=1mm, shadow=true](0.17808, 11.32075)(0.94521, 19.24528)
\rput[l](0.39726, 18.11321){1 km/h}
\psline[linecolor=red, linestyle=solid, linewidth=0.3mm](0.23288, 18.11321)(0.34247, 18.11321)
\rput[l](0.39726, 16.22642){10 km/h}
\psline[linecolor=green, linestyle=solid, linewidth=0.3mm](0.23288, 16.22642)(0.34247, 16.22642)
\rput[l](0.39726, 14.33962){100 km/h}
\psline[linecolor=blue, linestyle=solid, linewidth=0.3mm](0.23288, 14.33962)(0.34247, 14.33962)
\rput[l](0.39726, 12.45283){succ. CSI fb.}
\psline[linecolor=black, linestyle=dashed, linewidth=0.3mm](0.23288, 12.45283)(0.34247, 12.45283)
\end{pspicture}
\endgroup
 
\label{f:params_vs_uldlratio_nd}} \hfil \subfigure[Optimal pilot density.]{\begingroup
\unitlength=1mm
\psset{xunit=36.50000mm, yunit=353.33333mm, linewidth=0.1mm}
\psset{arrowsize=2pt 3, arrowlength=1.4, arrowinset=.4}\psset{axesstyle=frame}
\begin{pspicture}(-1.27397, -0.02830)(1.00000, 0.15000)
\psaxes[subticks=10, xlogBase=10, logLines=x, labels=all, xsubticks=1, ysubticks=1, Ox=-1, Oy=0, Dx=1, Dy=0.05]{-}(-1.00000, 0.00000)(-1.00000, 0.00000)(1.00020, 0.15020)%
\multips(-1.00000, 0.05000)(0, 0.05000){2}{\psline[linecolor=black, linestyle=dotted, linewidth=0.2mm](0, 0)(2.00000, 0)}
\rput[b](0.00000, -0.02830){Desired UL/DL ratio}
\psclip{\psframe(-1.00000, 0.00000)(1.00000, 0.15000)}
\psline[linecolor=red, plotstyle=curve, linestyle=solid, dotscale=1.5 1.5, showpoints=false, linewidth=0.3mm](-1.00000, 0.01667)(-0.77815, 0.01667)(-0.60206, 0.01667)(-0.47712, 0.01667)(-0.30103, 0.01667)(0.00000, 0.01667)(0.30103, 0.01667)(0.47712, 0.01667)(0.60206, 0.01667)(0.77815, 0.01667)(1.00000, 0.01667)
\psline[linecolor=red, plotstyle=curve, dotstyle=triangle, dotangle=0, linestyle=none, dotscale=1.5 1.5, showpoints=true, linewidth=0.3mm](-1.00000, 0.01667)(0.00000, 0.01667)(1.00000, 0.01667)
\psline[linecolor=red, plotstyle=curve, linestyle=dashed, dotscale=1.5 1.5, showpoints=false, linewidth=0.3mm](-1.00000, 0.01667)(-0.77815, 0.01667)(-0.60206, 0.01667)(-0.47712, 0.01667)(-0.30103, 0.01667)(0.00000, 0.01667)(0.30103, 0.01667)(0.47712, 0.01667)(0.60206, 0.01667)(0.77815, 0.01667)(1.00000, 0.01667)
\psline[linecolor=red, plotstyle=curve, dotstyle=triangle, dotangle=0, linestyle=none, dotscale=1.5 1.5, showpoints=true, linewidth=0.3mm](-1.00000, 0.01667)(0.00000, 0.01667)(1.00000, 0.01667)
\psline[linecolor=red, plotstyle=curve, linestyle=solid, dotscale=1.5 1.5, showpoints=false, linewidth=0.3mm](-1.00000, 0.01667)(-0.77815, 0.01667)(-0.60206, 0.01667)(-0.47712, 0.01667)(-0.30103, 0.01667)(0.00000, 0.01667)(0.30103, 0.02104)(0.47712, 0.04229)(0.60206, 0.06271)(0.77815, 0.06937)(1.00000, 0.06937)
\psline[linecolor=red, plotstyle=curve, dotstyle=triangle, dotangle=180, linestyle=none, dotscale=1.5 1.5, showpoints=true, linewidth=0.3mm](-1.00000, 0.01667)(0.00000, 0.01667)(1.00000, 0.06937)
\psline[linecolor=red, plotstyle=curve, linestyle=dashed, dotscale=1.5 1.5, showpoints=false, linewidth=0.3mm](-1.00000, 0.01667)(-0.77815, 0.01667)(-0.60206, 0.01667)(-0.47712, 0.01667)(-0.30103, 0.01667)(0.00000, 0.01667)(0.30103, 0.01729)(0.47712, 0.01792)(0.60206, 0.02354)(0.77815, 0.05313)(1.00000, 0.06937)
\psline[linecolor=red, plotstyle=curve, dotstyle=triangle, dotangle=180, linestyle=none, dotscale=1.5 1.5, showpoints=true, linewidth=0.3mm](-1.00000, 0.01667)(0.00000, 0.01667)(1.00000, 0.06937)
\psline[linecolor=green, plotstyle=curve, linestyle=solid, dotscale=1.5 1.5, showpoints=false, linewidth=0.3mm](-1.00000, 0.01667)(-0.77815, 0.01667)(-0.60206, 0.01667)(-0.47712, 0.01667)(-0.30103, 0.01667)(0.00000, 0.01667)(0.30103, 0.01667)(0.47712, 0.01667)(0.60206, 0.01667)(0.77815, 0.01667)(1.00000, 0.01667)
\psline[linecolor=green, plotstyle=curve, dotstyle=triangle, dotangle=0, linestyle=none, dotscale=1.5 1.5, showpoints=true, linewidth=0.3mm](-1.00000, 0.01667)(0.00000, 0.01667)(1.00000, 0.01667)
\psline[linecolor=green, plotstyle=curve, linestyle=dashed, dotscale=1.5 1.5, showpoints=false, linewidth=0.3mm](-1.00000, 0.01667)(-0.77815, 0.01667)(-0.60206, 0.01667)(-0.47712, 0.01667)(-0.30103, 0.01667)(0.00000, 0.01667)(0.30103, 0.01667)(0.47712, 0.01667)(0.60206, 0.01667)(0.77815, 0.01667)(1.00000, 0.01667)
\psline[linecolor=green, plotstyle=curve, dotstyle=triangle, dotangle=0, linestyle=none, dotscale=1.5 1.5, showpoints=true, linewidth=0.3mm](-1.00000, 0.01667)(0.00000, 0.01667)(1.00000, 0.01667)
\psline[linecolor=green, plotstyle=curve, linestyle=solid, dotscale=1.5 1.5, showpoints=false, linewidth=0.3mm](-1.00000, 0.01667)(-0.77815, 0.01667)(-0.60206, 0.01667)(-0.47712, 0.01667)(-0.30103, 0.01667)(0.00000, 0.01792)(0.30103, 0.03167)(0.47712, 0.05604)(0.60206, 0.06646)(0.77815, 0.06917)(1.00000, 0.06917)
\psline[linecolor=green, plotstyle=curve, dotstyle=triangle, dotangle=180, linestyle=none, dotscale=1.5 1.5, showpoints=true, linewidth=0.3mm](-1.00000, 0.01667)(0.00000, 0.01792)(1.00000, 0.06917)
\psline[linecolor=green, plotstyle=curve, linestyle=dashed, dotscale=1.5 1.5, showpoints=false, linewidth=0.3mm](-1.00000, 0.01667)(-0.77815, 0.01667)(-0.60206, 0.01667)(-0.47712, 0.01667)(-0.30103, 0.01667)(0.00000, 0.01729)(0.30103, 0.01958)(0.47712, 0.02667)(0.60206, 0.04271)(0.77815, 0.06521)(1.00000, 0.06917)
\psline[linecolor=green, plotstyle=curve, dotstyle=triangle, dotangle=180, linestyle=none, dotscale=1.5 1.5, showpoints=true, linewidth=0.3mm](-1.00000, 0.01667)(0.00000, 0.01729)(1.00000, 0.06917)
\psline[linecolor=blue, plotstyle=curve, linestyle=solid, dotscale=1.5 1.5, showpoints=false, linewidth=0.3mm](-1.00000, 0.01667)(-0.77815, 0.01667)(-0.60206, 0.01667)(-0.47712, 0.01667)(-0.30103, 0.01667)(0.00000, 0.01667)(0.30103, 0.01667)(0.47712, 0.01667)(0.60206, 0.01667)(0.77815, 0.01667)(1.00000, 0.01667)
\psline[linecolor=blue, plotstyle=curve, dotstyle=triangle, dotangle=0, linestyle=none, dotscale=1.5 1.5, showpoints=true, linewidth=0.3mm](-1.00000, 0.01667)(0.00000, 0.01667)(1.00000, 0.01667)
\psline[linecolor=blue, plotstyle=curve, linestyle=dashed, dotscale=1.5 1.5, showpoints=false, linewidth=0.3mm](-1.00000, 0.01667)(-0.77815, 0.01667)(-0.60206, 0.01667)(-0.47712, 0.01667)(-0.30103, 0.01667)(0.00000, 0.01667)(0.30103, 0.01667)(0.47712, 0.01667)(0.60206, 0.01667)(0.77815, 0.01667)(1.00000, 0.01667)
\psline[linecolor=blue, plotstyle=curve, dotstyle=triangle, dotangle=0, linestyle=none, dotscale=1.5 1.5, showpoints=true, linewidth=0.3mm](-1.00000, 0.01667)(0.00000, 0.01667)(1.00000, 0.01667)
\psline[linecolor=blue, plotstyle=curve, linestyle=solid, dotscale=1.5 1.5, showpoints=false, linewidth=0.3mm](-1.00000, 0.09292)(-0.77815, 0.09292)(-0.60206, 0.09375)(-0.47712, 0.09542)(-0.30103, 0.09542)(0.00000, 0.09542)(0.30103, 0.09542)(0.47712, 0.09542)(0.60206, 0.09542)(0.77815, 0.09542)(1.00000, 0.09542)
\psline[linecolor=blue, plotstyle=curve, dotstyle=triangle, dotangle=180, linestyle=none, dotscale=1.5 1.5, showpoints=true, linewidth=0.3mm](-1.00000, 0.09292)(0.00000, 0.09542)(1.00000, 0.09542)
\psline[linecolor=blue, plotstyle=curve, linestyle=dashed, dotscale=1.5 1.5, showpoints=false, linewidth=0.3mm](-1.00000, 0.09292)(-0.77815, 0.09292)(-0.60206, 0.09375)(-0.47712, 0.09542)(-0.30103, 0.09542)(0.00000, 0.09542)(0.30103, 0.09542)(0.47712, 0.09542)(0.60206, 0.09542)(0.77815, 0.09542)(1.00000, 0.09542)
\psline[linecolor=blue, plotstyle=curve, dotstyle=triangle, dotangle=180, linestyle=none, dotscale=1.5 1.5, showpoints=true, linewidth=0.3mm](-1.00000, 0.09292)(0.00000, 0.09542)(1.00000, 0.09542)
\rput(0.80000,0.04000){\cnodeput[fillstyle=solid, fillcolor=lightgray](0,0){xx}{\scalebox{0.7}{I}}}
\endpsclip
\rput[t]{90}(-1.27397, 0.07500){opt. pilot density}
\psframe[linecolor=black, fillstyle=solid, fillcolor=white, shadowcolor=lightgray, shadowsize=1mm, shadow=true](-0.94521, 0.07075)(-0.31507, 0.14434)
\rput[l](-0.72603, 0.13585){1 km/h}
\psline[linecolor=red, linestyle=solid, linewidth=0.3mm](-0.89041, 0.13585)(-0.78082, 0.13585)
\rput[l](-0.72603, 0.12170){10 km/h}
\psline[linecolor=green, linestyle=solid, linewidth=0.3mm](-0.89041, 0.12170)(-0.78082, 0.12170)
\rput[l](-0.72603, 0.10755){100 km/h}
\psline[linecolor=blue, linestyle=solid, linewidth=0.3mm](-0.89041, 0.10755)(-0.78082, 0.10755)
\rput[l](-0.72603, 0.09340){uplink}
\psline[linecolor=black, linestyle=solid, linewidth=0.3mm](-0.89041, 0.09340)(-0.78082, 0.09340)
\psline[linecolor=black, linestyle=solid, linewidth=0.3mm](-0.89041, 0.09340)(-0.78082, 0.09340)
\psdots[linecolor=black, linestyle=solid, linewidth=0.3mm, dotstyle=triangle, dotangle=0, dotscale=1.5 1.5, linecolor=black](-0.83562, 0.09340)
\rput[l](-0.72603, 0.07925){downlink}
\psline[linecolor=black, linestyle=solid, linewidth=0.3mm](-0.89041, 0.07925)(-0.78082, 0.07925)
\psline[linecolor=black, linestyle=solid, linewidth=0.3mm](-0.89041, 0.07925)(-0.78082, 0.07925)
\psdots[linecolor=black, linestyle=solid, linewidth=0.3mm, dotstyle=triangle, dotangle=180, dotscale=1.5 1.5, linecolor=black](-0.83562, 0.07925)

\end{pspicture}
\endgroup
 
\label{f:params_vs_uldlratio_pd}}}
\caption{Optimal choice of parameters as a function of the desired UL/DL ratio.} \label{f:params_vs_uldlratio}
\end{figure*}

\begin{figure*}
\centerline{\subfigure{\begingroup
\unitlength=1mm
\psset{xunit=29.20000mm, yunit=2.12000mm, linewidth=0.1mm}
\psset{arrowsize=2pt 3, arrowlength=1.4, arrowinset=.4}\psset{axesstyle=frame}
\begin{pspicture}(-0.34247, -4.71698)(2.50000, 25.00000)
\psaxes[subticks=10, xlogBase=10, logLines=x, labels=all, xsubticks=1, ysubticks=1, Ox=0, Oy=0, Dx=1, Dy=5]{-}(0.00000, 0.00000)(0.00000, 0.00000)(2.50020, 25.00020)%
\multips(0.00000, 5.00000)(0, 5.00000){4}{\psline[linecolor=black, linestyle=dotted, linewidth=0.2mm](0, 0)(2.50000, 0)}
\rput[b](1.25000, -4.71698){UE speed [km/h]}
\psclip{\psframe(0.00000, 0.00000)(2.50000, 25.00000)}
\psline[linecolor=red, plotstyle=curve, linestyle=solid, dotscale=1.5 1.5, showpoints=false, linewidth=0.3mm](0.00000, 4.44771)(0.25001, 4.41446)(0.50000, 4.31280)(0.75000, 4.02062)(1.00000, 3.22305)(1.25000, 1.28281)(1.50000, 1.07262)(1.75000, 6.82178)(2.00000, 13.52312)(2.25000, 21.04543)(2.50000, 166.82709)
\psline[linecolor=red, plotstyle=curve, linestyle=none, dotscale=1.5 1.5, showpoints=false, linewidth=0.3mm](0.00000, 4.44771)(1.00000, 3.22305)(2.00000, 13.52312)
\psline[linecolor=red, plotstyle=curve, linestyle=dashed, dotscale=1.5 1.5, showpoints=false, linewidth=0.3mm](0.00000, 0.00000)(0.25001, 0.00000)(0.50000, 0.00000)(0.75000, 0.00000)(1.00000, 0.00880)(1.25000, 0.12705)(1.50000, 0.90140)(1.75000, 6.63311)(2.00000, 13.55589)(2.25000, 21.03753)(2.50000, 166.74436)
\psline[linecolor=red, plotstyle=curve, linestyle=none, dotscale=1.5 1.5, showpoints=false, linewidth=0.3mm](0.00000, 0.00000)(1.00000, 0.00880)(2.00000, 13.55589)
\psline[linecolor=green, plotstyle=curve, linestyle=solid, dotscale=1.5 1.5, showpoints=false, linewidth=0.3mm](0.00000, 0.26435)(0.25001, 0.25215)(0.50000, 0.21468)(0.75000, 0.12411)(1.00000, 0.09324)(1.25000, 1.23481)(1.50000, 5.02220)(1.75000, 10.91189)(2.00000, 14.38210)(2.25000, 17.38696)(2.50000, 111.46950)
\psline[linecolor=green, plotstyle=curve, linestyle=none, dotscale=1.5 1.5, showpoints=false, linewidth=0.3mm](0.00000, 0.26435)(1.00000, 0.09324)(2.00000, 14.38210)
\psline[linecolor=green, plotstyle=curve, linestyle=dashed, dotscale=1.5 1.5, showpoints=false, linewidth=0.3mm](0.00000, 1.47265)(0.25001, 1.48305)(0.50000, 1.51466)(0.75000, 1.60194)(1.00000, 1.56070)(1.25000, 2.38327)(1.50000, 5.05822)(1.75000, 10.82480)(2.00000, 14.39774)(2.25000, 17.38531)(2.50000, 111.45614)
\psline[linecolor=green, plotstyle=curve, linestyle=none, dotscale=1.5 1.5, showpoints=false, linewidth=0.3mm](0.00000, 1.47265)(1.00000, 1.56070)(2.00000, 14.39774)
\psline[linecolor=blue, plotstyle=curve, linestyle=solid, dotscale=1.5 1.5, showpoints=false, linewidth=0.3mm](0.00000, 6.66585)(0.25001, 6.70491)(0.50000, 6.82647)(0.75000, 7.19260)(1.00000, 8.18782)(1.25000, 10.12291)(1.50000, 11.98408)(1.75000, 13.76643)(2.00000, 14.81660)(2.25000, 16.13155)(2.50000, 96.25063)
\psline[linecolor=blue, plotstyle=curve, linestyle=none, dotscale=1.5 1.5, showpoints=false, linewidth=0.3mm](0.00000, 6.66585)(1.00000, 8.18782)(2.00000, 14.81660)
\psline[linecolor=blue, plotstyle=curve, linestyle=dashed, dotscale=1.5 1.5, showpoints=false, linewidth=0.3mm](0.00000, 6.04081)(0.25001, 6.07687)(0.50000, 6.19328)(0.75000, 6.56059)(1.00000, 7.61790)(1.25000, 9.85420)(1.50000, 11.92501)(1.75000, 13.74126)(2.00000, 14.81953)(2.25000, 16.13126)(2.50000, 96.24843)
\psline[linecolor=blue, plotstyle=curve, linestyle=none, dotscale=1.5 1.5, showpoints=false, linewidth=0.3mm](0.00000, 6.04081)(1.00000, 7.61790)(2.00000, 14.81953)
\rput(0.50000,5.00000){\cnodeput[fillstyle=solid, fillcolor=lightgray](0,0){xx}{\scalebox{0.7}{J}}}
\rput(2.20000,13.00000){\cnodeput[fillstyle=solid, fillcolor=lightgray](0,0){xx}{\scalebox{0.7}{K}}}
\endpsclip
\rput[t]{90}(-0.34247, 12.50000){Rate gain through adaptation [\%]}
\psframe[linecolor=black, fillstyle=solid, fillcolor=white, shadowcolor=lightgray, shadowsize=1mm, shadow=true](0.06849, 14.15094)(1.02466, 24.05660)
\rput[l](0.34247, 22.64151){UL/DL 1:6}
\psline[linecolor=red, linestyle=solid, linewidth=0.3mm](0.13699, 22.64151)(0.27397, 22.64151)
\rput[l](0.34247, 20.28302){UL/DL 1:1}
\psline[linecolor=green, linestyle=solid, linewidth=0.3mm](0.13699, 20.28302)(0.27397, 20.28302)
\rput[l](0.34247, 17.92453){UL/DL 6:1}
\psline[linecolor=blue, linestyle=solid, linewidth=0.3mm](0.13699, 17.92453)(0.27397, 17.92453)
\rput[l](0.34247, 15.56604){succ. CSI fb.}
\psline[linecolor=black, linestyle=dashed, linewidth=0.3mm](0.13699, 15.56604)(0.27397, 15.56604)
\end{pspicture}
\endgroup
 
\label{f:gain_vs_speed}} \hfil
\subfigure{\begingroup
\unitlength=1mm
\psset{xunit=36.50000mm, yunit=2.12000mm, linewidth=0.1mm}
\psset{arrowsize=2pt 3, arrowlength=1.4, arrowinset=.4}\psset{axesstyle=frame}
\begin{pspicture}(-1.27397, -4.71698)(1.00000, 25.00000)
\psaxes[subticks=10, xlogBase=10, logLines=x, labels=all, xsubticks=1, ysubticks=1, Ox=-1, Oy=0, Dx=1, Dy=5]{-}(-1.00000, 0.00000)(-1.00000, 0.00000)(1.00020, 25.00020)%
\multips(-1.00000, 5.00000)(0, 5.00000){4}{\psline[linecolor=black, linestyle=dotted, linewidth=0.2mm](0, 0)(2.00000, 0)}
\rput[b](0.00000, -4.71698){Desired UL/DL ratio}
\psclip{\psframe(-1.00000, 0.00000)(1.00000, 25.00000)}
\psline[linecolor=red, plotstyle=curve, linestyle=solid, dotscale=1.5 1.5, showpoints=false, linewidth=0.3mm](-1.00000, 5.59908)(-0.77815, 4.44771)(-0.60206, 3.42667)(-0.47712, 2.61341)(-0.30103, 1.59075)(0.00000, 0.26435)(0.30103, 0.18009)(0.47712, 1.53945)(0.60206, 3.64936)(0.77815, 6.66585)(1.00000, 9.38414)
\psline[linecolor=red, plotstyle=curve, linestyle=none, dotscale=1.5 1.5, showpoints=false, linewidth=0.3mm](-1.00000, 5.59908)(0.00000, 0.26435)(1.00000, 9.38414)
\psline[linecolor=red, plotstyle=curve, linestyle=dashed, dotscale=1.5 1.5, showpoints=false, linewidth=0.3mm](-1.00000, 0.06119)(-0.77815, 0.00000)(-0.60206, 0.00077)(-0.47712, 0.14150)(-0.30103, 0.43548)(0.00000, 1.47265)(0.30103, 2.63274)(0.47712, 3.43685)(0.60206, 4.31935)(0.77815, 6.04081)(1.00000, 8.86016)
\psline[linecolor=red, plotstyle=curve, linestyle=none, dotscale=1.5 1.5, showpoints=false, linewidth=0.3mm](-1.00000, 0.06119)(0.00000, 1.47265)(1.00000, 8.86016)
\psline[linecolor=green, plotstyle=curve, linestyle=solid, dotscale=1.5 1.5, showpoints=false, linewidth=0.3mm](-1.00000, 4.25273)(-0.77815, 3.22305)(-0.60206, 2.27796)(-0.47712, 1.66283)(-0.30103, 0.81372)(0.00000, 0.09324)(0.30103, 1.16295)(0.47712, 3.61200)(0.60206, 5.68507)(0.77815, 8.18782)(1.00000, 10.37050)
\psline[linecolor=green, plotstyle=curve, linestyle=none, dotscale=1.5 1.5, showpoints=false, linewidth=0.3mm](-1.00000, 4.25273)(0.00000, 0.09324)(1.00000, 10.37050)
\psline[linecolor=green, plotstyle=curve, linestyle=dashed, dotscale=1.5 1.5, showpoints=false, linewidth=0.3mm](-1.00000, 0.02137)(-0.77815, 0.00880)(-0.60206, 0.07667)(-0.47712, 0.27831)(-0.30103, 0.59419)(0.00000, 1.56070)(0.30103, 2.94296)(0.47712, 4.29728)(0.60206, 5.43672)(0.77815, 7.61790)(1.00000, 9.99147)
\psline[linecolor=green, plotstyle=curve, linestyle=none, dotscale=1.5 1.5, showpoints=false, linewidth=0.3mm](-1.00000, 0.02137)(0.00000, 1.56070)(1.00000, 9.99147)
\psline[linecolor=blue, plotstyle=curve, linestyle=solid, dotscale=1.5 1.5, showpoints=false, linewidth=0.3mm](-1.00000, 13.43220)(-0.77815, 13.52312)(-0.60206, 13.64199)(-0.47712, 13.78219)(-0.30103, 14.03050)(0.00000, 14.38210)(0.30103, 14.62130)(0.47712, 14.71470)(0.60206, 14.76451)(0.77815, 14.81660)(1.00000, 14.86003)
\psline[linecolor=blue, plotstyle=curve, linestyle=none, dotscale=1.5 1.5, showpoints=false, linewidth=0.3mm](-1.00000, 13.43220)(0.00000, 14.38210)(1.00000, 14.86003)
\psline[linecolor=blue, plotstyle=curve, linestyle=dashed, dotscale=1.5 1.5, showpoints=false, linewidth=0.3mm](-1.00000, 13.48923)(-0.77815, 13.55589)(-0.60206, 13.67034)(-0.47712, 13.81920)(-0.30103, 14.05809)(0.00000, 14.39774)(0.30103, 14.62967)(0.47712, 14.72042)(0.60206, 14.76885)(0.77815, 14.81953)(1.00000, 14.86181)
\psline[linecolor=blue, plotstyle=curve, linestyle=none, dotscale=1.5 1.5, showpoints=false, linewidth=0.3mm](-1.00000, 13.48923)(0.00000, 14.39774)(1.00000, 14.86181)
\endpsclip
\rput[t]{90}(-1.27397, 12.50000){Rate gain through adaptation [\%]}
\psframe[linecolor=black, fillstyle=solid, fillcolor=white, shadowcolor=lightgray, shadowsize=1mm, shadow=true](-0.94521, 14.15094)(-0.18507, 24.05660)
\rput[l](-0.72603, 22.64151){1 km/h}
\psline[linecolor=red, linestyle=solid, linewidth=0.3mm](-0.89041, 22.64151)(-0.78082, 22.64151)
\rput[l](-0.72603, 20.28302){10 km/h}
\psline[linecolor=green, linestyle=solid, linewidth=0.3mm](-0.89041, 20.28302)(-0.78082, 20.28302)
\rput[l](-0.72603, 17.92453){100 km/h}
\psline[linecolor=blue, linestyle=solid, linewidth=0.3mm](-0.89041, 17.92453)(-0.78082, 17.92453)
\rput[l](-0.72603, 15.56604){succ. CSI fb.}
\psline[linecolor=black, linestyle=dashed, linewidth=0.3mm](-0.89041, 15.56604)(-0.78082, 15.56604)
\end{pspicture}
\endgroup
 
\label{f:gain_vs_uldlratio}}}
\caption{Weighted joint UL/DL sum-rate gains through adaptation (as opposed to fixed parameters).} \label{f:gain}
\end{figure*}

Fig.~\ref{f:params_vs_speed} shows the dependency of the terminal velocity and the optimal number of CSI feedback bits $\Nb$ per spatial channel coefficient and PRB (plot~\ref{f:params_vs_speed_nd}) as well as the optimal pilot densities in \ac{UL} and \ac{DL} (plot~\ref{f:params_vs_speed_pd}).
    For low speeds, and especially if the DL is considered important, it is beneficial to invest large extents of \ac{UL} capacity into CSI feedback~~\cnodeput(0,+1mm){t}{\scalebox{0.5}{A}}~~. The difference between non-succ. and succ. CSI feedback is rather small, as the weighted sum-rate optimization makes the system invest the gained capacity into boosting the DL, rather than decreasing CSI feedback. For moderate speeds, there is little difference between non-succ. and succ. CSI feedback, until for large speeds, DL performance is so strongly impaired through CSI feedback delay that the optimum extent of CSI feedback decreases until the system uses TDM in the DL~~\cnodeput(0,+1mm){t}{\scalebox{0.5}{B}}~~. If strong priority is given to the \ac{UL}, the system only invests into succ. CSI feedback and operates in TDM otherwise~~\cnodeput(0,+1mm){t}{\scalebox{0.5}{C}}~~. Plot~\ref{f:params_vs_speed_pd} shows that the \ac{UL} pilot density remains constant except for large UT speeds~~\cnodeput(0,+1mm){t}{\scalebox{0.5}{D}}~~. Depending on the desired \ac{UL}/\ac{DL} ratio, the DL pilot density switches between two modes: A low density for spatial multiplexing operation and a higher density for TDM operation~~\cnodeput(0,+1mm){t}{\scalebox{0.5}{E}}~~. For large UT speeds, CSI feedback delay becomes the dominant issue, such that pilot density in the DL is reduced again~~\cnodeput(0,+1mm){t}{\scalebox{0.5}{F}}~~.
    
    \medskip
    
    Fig.~\ref{f:params_vs_uldlratio} shows the same parameters, but as a function of target \ac{UL}/\ac{DL} ratio. We can see a similar trend as before, hence when the focus is shifted more towards the \ac{UL}, and for larger UT speeds, less \ac{UL} is invested into CSI feedback~~\cnodeput(0,+1mm){t}{\scalebox{0.5}{G}}~~. As succ. CSI feedback improves the performance/feedback ratio, it leads to the fact that even for a strong focus on the \ac{UL} it is still beneficial to operate the DL in spatial multiplexing mode~~\cnodeput(0,+1mm){t}{\scalebox{0.5}{H}}~~. If the DL is operated in TDM, it is then beneficial to increase the DL pilot density, as only one pilot sequence is needed instead of $\NBS + K$ as for spatial multiplexing.
    
    \medskip
    
    Fig.~\ref{f:gain} finally shows the weighted \ac{UL}/\ac{DL} sum rate gain that can be achieved through an adaptive usage of pilot densities and CSI feedback quantity, as opposed to fixed parameters $\Nb=12$, $\rho_{\textnormal{UL}}=\rho_{\textnormal{DL}}=0.017$. These parameters have shown to be optimal on average for a terminal speed of $v=10$ km/h and a desired \ac{UL}/\ac{DL}-ratio of $1:1$. In both cases, we have the option of switching between spatial multiplexing and TDM in the DL. Gains are shown as a function of UT speed in plot~\ref{f:gain_vs_speed} and of the desired \ac{UL}/\ac{DL} ratio in plot~\ref{f:gain_vs_uldlratio}, respectively. In regimes of low speed, a gain in the order of a few percent is visible, as an increase in CSI feedback quantity beyond $\Nb=12$ can still improve rates~~\cnodeput(0,+1mm){t}{\scalebox{0.5}{I}}~~. A large gain of adaptation is visible for large UT velocities, as here both \ac{UL} and \ac{DL} ask for more dense pilot structures, in particular in conjunction with DL TDM. Plot~\ref{f:gain_vs_uldlratio} shows adaptation gains as soon as a different target \ac{UL}/\ac{DL} ratio is desired. In general, adaptation gains are reduced if succ. CSI feedback is employed, as this requires less sacrifice of \ac{UL} rates. In practical systems, we expect the gains through adaptation to be larger, as both channel estimation and CSI feedback schemes will perform significantly worse than the information theoretical bounds observed here, such that it will be even more essential to optimize the trade-off between \ac{UL} and \ac{DL} rates. We expect an additional gain if in the \ac{DL}, pilot densities specific for BS antennas and stream-specific can be adjusted individually.  

  \section{Conclusions}\label{sec:conclusions}
    \begin{table}
      \begin{tabular}{p{0.2cm} c|c c}
	$N_b$ & ~ & \multicolumn{2}{c}{Velocity / Large Scale Scattering} \\
	\multicolumn{2}{l|}{~} & \multicolumn{1}{c|}{low} & high \\ \hline\hline
	\multirow{2}{*}{\rotateleft{UL/DL ratio~ ~ ~ ~}} & $<1$ & 
	  \begin{minipage}{2.8cm}
	    ~\\\textbf{high} $\Nb$ -- a major part of the \ac{UL} is used to improve the \ac{DL} throughput, as the \ac{UL} is less important and the high coherence facilitates precise CSIT.\\
	  \end{minipage}\rput(0, 0){\psline[arrowsize=3pt 4, arrowlength=1.4, arrowinset=.4, linecolor=darkred]{->}(0.2cm, -4.3cm)(0.2cm, 1.20cm)}%
	  \rput(0, 0){\psline[arrowsize=3pt 4, arrowlength=1.4, arrowinset=.4, linecolor=darkred]{<-}(-2.8cm, -1.2cm)(4cm, -1.2cm)}%
	  \rput(0.2cm, -1.2cm){\psframebox[fillcolor=white, fillstyle=solid, linestyle=none]{\textcolor{darkred}{Increasing $N_b$}}}%
	  &
	  \begin{minipage}{3.5cm}
	    ~\\\textbf{moderate} $\Nb$ -- only a moderate part of the \ac{UL} is spent for CSI feedback as the low coherence
	    implies a feedback-delay-noise dominating channel estimation noise.\\
	  \end{minipage}\\ \cline{2-2}
	~ & $>1$ & 
	  \begin{minipage}{2.8cm}
	    \textbf{moderate} $\Nb$ -- the \ac{UL} obtains more priority and therefore cannot trade the same number of feedback bits
	    to improve the DL as in case of a lower \ac{UL}/\ac{DL} ratio.
	  \end{minipage} &
	  \begin{minipage}{3.5cm}
	    ~\\~\\\textbf{no CSI feedb.} -- the prioritized \ac{UL} cannot give up enough resources for feedback and
	    the high velocity implies a dominant feedback-delay noise. It is best to operate the DL in TDM mode and abandon BS precoding.\\
	  \end{minipage} \\
      \end{tabular}
      \caption{Qualitative summary of the optimal number of feedback bits depending on the \ac{UL}/\ac{DL} ratio and the channel
      coherence.}
      \label{fig:results:qualitative.summary}
    \end{table}
    Our analysis revealed two apparent trends, which are summarized in Table \ref{fig:results:qualitative.summary}:
    CSI feedback becomes more beneficial for decreasing UT speed (and less scattering) as well as for an increasing weight on the \ac{DL} rate. 
    Hence, a system might use a multi-cross layer approach in order to adaptively control the physical layer (pilot structure and CSI feedback)
    depending on the application (\ac{UL}/\ac{DL} ratio) as well as depending on the channel (velocity, scattering). 
    In addition, our analysis demonstrated the potential of succ. CSI feedback to scale down the signaling overhead in regimes of low to moderate terminal velocity. Our multi-cross-layer approach could be further extended to include QoS constraints such as latency and packet error rate as well as more degrees of freedom, e.\,g., the change of physical layer parameters such as the block size in time and frequency, which would imply significant changes for the architecture of mobile communication systems.

  %
  %
  \bibliographystyle{IEEEtran}
  \bibliography{IEEEabrv,my-references,figures/General_Bibliography}
\end{document}